\documentclass[manuscript]{acmart}
\AtBeginDocument{%
  }

\setcopyright{acmlicensed}
\copyrightyear{2018}
\acmYear{2018}
\acmDOI{XXXXXXX.XXXXXXX}

\acmConference[Conference acronym 'XX]{Make sure to enter the correct
  conference title from your rights confirmation emai}{June 03--05,
  2018}{Woodstock, NY}
\acmISBN{978-1-4503-XXXX-X/18/06}

\acmSubmissionID{7595}

\usepackage{algorithm}
\usepackage{algorithmic}
\usepackage{xcolor} 
\usepackage{caption}
\usepackage{graphicx, subfig}

\usepackage{amssymb}
\usepackage{multirow}
\usepackage{array}
\usepackage{amsmath}
\usepackage{booktabs}
\usepackage{makecell}
\usepackage{url}
\usepackage{float}

\newcommand{\texticon}[1]{%
  \raisebox{-0.3\height}{%
    \includegraphics[height=1.1em]{#1}%
  }%
}

\newcommand{\texticonnn}[1]{%
  \raisebox{-0.5\height}{%
    \includegraphics[height=1.6em]{#1}%
  }%
}

\newcommand{\textlab}[1]{%
  \raisebox{-0.3\height}{%
    \includegraphics[height=3em]{#1}%
  }%
}

\newcommand{\textagent}[1]{%
  \raisebox{-0.1\height}{%
    \includegraphics[height=1.1em]{#1}%
  }%
}

\newcommand{\interfaceicon}[1]{%
  \raisebox{-0.08\height}{%
    \includegraphics[height=0.9em]{#1}%
  }%
}

\begin{document}

\title{Facilitating Video Story Interaction with Multi-Agent Collaborative System}

\author{Yiwen Zhang}
\email{yzhang452@connect.hkust-gz.edu.cn}
\affiliation{%
 \institution{The Hong Kong University of Science and Technology (Guangzhou)}
 \country{China}
}

\author{Jianing Hao}
\affiliation{%
 \institution{The Hong Kong University of Science and Technology (Guangzhou)}
 \country{China}
}

\author{Zhan Wang}
\affiliation{%
 \institution{The Hong Kong University of Science and Technology (Guangzhou)}
 \country{China}
}

\author{Hongling Sheng}
\affiliation{%
 \institution{The Hong Kong University of Science and Technology (Guangzhou)}
 \country{China}
}

\author{Wei Zeng}
\affiliation{%
 \institution{The Hong Kong University of Science and Technology (Guangzhou)}
 \country{China}
}

\renewcommand{\shortauthors}{Zhang et al.}

\begin{abstract}
  

Video story interaction enables viewers to engage with and explore narrative content for personalized experiences. 
However, existing methods are limited to user selection, specially designed narratives, and lack customization. 
To address this, we propose an interactive system based on user intent. Our system uses a Vision Language Model (VLM) to enable machines to understand video stories, combining Retrieval-Augmented Generation (RAG) and a Multi-Agent System (MAS) to create evolving characters and scene experiences.
It includes three stages: 1) Video story processing, utilizing VLM and prior knowledge to simulate human understanding of stories across three modalities. 2) Multi-space chat, creating growth-oriented characters through MAS interactions based on user queries and story stages. 3) Scene customization, expanding and visualizing various story scenes mentioned in dialogue.
Applied to the Harry Potter series, our study shows the system effectively portrays emergent character social behavior and growth, enhancing the interactive experience in the video story world.

\end{abstract}




\maketitle

\section{Introduction}


In an age where digital storytelling \cite{n22} transcends mere observation, the quest for a deeper, participatory journey through video narratives becomes increasingly demanding \cite{n19}.
Users are not just satisfied with passively receiving content but want to actively experience and create it \cite{n18}.
Unlike TV series, interactive video story \cite{n19,n18} allow audience input to shape their experience.
Movies that follow this concept, such as Netflix's Black Mirror: Bandersnatch \cite{n20} and CtrlMovie's Late Shift \cite{n21}, have gained popularity through streaming releases.
This narrative approach not only offers an immersive experience akin to traditional linear storytelling \cite{n26} but also empowers the audience to influence and alter the course of the story, thereby significantly enhancing their engagement and satisfaction.

While interactive video story has immense potential to enhance user engagement, it faces several challenges in practice. For one, this mode of interaction can disrupt the flow of the narrative and may not be suited for video stories with fixed plots like the ``Harry Potter'' series \cite{n16,n9,n7,n19}.
Secondly, existing interactive forms are confined to decision-making and lack a variety of interaction types that could enrich user experiences and emotional responses \cite{n16,n25,n19}.
Additionally, current methods fall short on deep personalization.
Although users can alter the story's trajectory by selecting different storylines, these narratives remain pre-defined by the creators and do not offer customization based on users' personal preferences and past interactions, making it difficult to achieve the desired level of tailored storytelling \cite{n18,n19,n15}.

Achieving the highly dynamic interactive narrative space \cite{n25} presents several challenges, including the seamless integration of narrative coherence and interactivity, all while designing for high levels of interaction and individualized customization without disrupting the narrative flow.
Firstly, machines currently struggle with the multi-modal comprehension of video stories and integrating this comprehension into the human-computer interaction (HCI) process \cite{v31}.
This difficulty somewhat confines the level of interactivity achievable with machine systems.
Secondly, the tools associated with the characters of user conversations \cite{v12,v22,n16} in the current research are less likely to achieve character growth over longer time spans \cite{m17} and generalizability to multiple characters.
This makes it difficult to produce interactive experiences with user-character dialog at low cost and easy-to-use \cite{v12,v22}.
Lastly, in visual storytelling, finding the right balance between interactivity and smooth narrative remains a challenge \cite{n19}, with limited technology available to enable users to smoothly transition between viewing the narrative and engaging interaction \cite{n25}.

To research and address the challenges mentioned above, we first conducted interviews with 12 professionals who have over five years of experience in three industries related to interacting with video stories (writing/directing, telling/hosting, and watching stories) to understand users' interaction intentions and needs.
Then, based on narrative theory, we designed a multi-modal understanding of video stories for machines and integrated this understanding into the users' interaction process.
To enrich the interactive experience, we designed multiple evolving characters to dialog with users.
For enhancing personalization, we incorporated user intents into three individualized plot options, making it easier for users to create their own unique story development.
We conducted a user study to comprehensively evaluate various aspects of our system's module functions, overall usability, interaction experience, and the growth potential of characters. 
We found that scene customization and characters with growth potential are most effective in enhancing users' learning and experience of video story content.

In summary, our main contributions consist of the following points:
\begin{itemize}
    \item A formative user study that identifies user needs and limitations in current video story interactions.
    \item Utilization of a Vision Language Model (VLM) to enable machines to understand video stories through vision, audio, and plot text modalities, simulating human comprehension.
    \item An approach integrating Retrieval Augmented Generation (RAG) with Multi-Agent System (MAS) to enhance user experience, allowing for dynamic dialogues with evolving characters and customizable narrative scenes.
\end{itemize}






\section{Background and Related Work}

\subsection{Video Story Understanding}


Video story understanding \cite{v31} enables machines to grasp the narrative information within a video story, such as the setting, characters, and plot. 
This not only provides humans with a broader perspective for understanding video stories \cite{v31,v33}, enriching the experience of interactive storytelling \cite{v37} but can also be applied in the domain of video story recommendation \cite{v35}.


Techniques for video story understanding mainly fall into two categories: rule-based and data-driven. 
Rule-based approaches manually analyze video content using narrative theories \cite{n12,n1,n3,n6,n14}. 
The advantage of this method lies in its strong interpretability, but it lacks flexibility and struggles to adapt to new situations that do not fully match predefined rules. 
On the other hand, data-driven approaches make use of existing video data to learn about characters \cite{v32}, structure \cite{v35}, and plot \cite{v31} patterns within video stories, which can then be applied to new videos \cite{v8}.
For example, RoleNet \cite{v32} constructs a social network of movie characters based on a social story segmentation method. 
Nan et al. \cite{v33} analyzed the status and influence of characters within the plot through deep learning.  
MovieGraphs \cite{v34} proposed a dataset describing character reaction graphs under video story contexts. 
DramaQA \cite{v31} trains machines to answer questions about video stories, stratifying video story understanding by difficulty.
Chen et al. \cite{v12} introduced a benchmark for matching agents with Harry Potter characters driven by LLM.
Character-LLM \cite{v22} utilizes character experience data to train agents that align with character traits.
Even though the capabilities of LLMs have enhanced the machine's understanding of characters, both \cite{v12,v22} cannot become universal tools due to their training data being confined to specific stories.
The FABULA analytical framework \cite{v15} integrates the capabilities of RAG \cite{t3} for the comprehension of story plots, which can be utilized not only for intelligence reporting but also for extracting information from highly narrative video storylines.
Data-driven methods enhance the transferability of video story understanding, but currently, such approaches are confined to understanding individual elements of video stories, making it challenging for machines to achieve an integrated understanding of multiple narrative elements.
Therefore, our goal is to use VLM and prior knowledge to understand video stories by integrating multiple narrative elements.

\subsection{LLM-Based Multi-Agent System}


Multi-Agent System (MAS) \cite{t4,m22,m23,m24,m25} refers to multiple agents completing tasks collaboratively, emphasizing diverse agent profiles \cite{m23,m36}, inter-agent interactions \cite{m12,m34}, and collective decision-making processes \cite{m21,m35}.
It affords the advantages of scalability, flexibility, and efficiency in creating opportunities for skill-based intelligent task allocation and garnering valuable feedback from a diversity of agent roles \cite{m18}.
MAS can facilitate discussion and resolution in simulations \cite{m9,m19}, game environments \cite{m6,m20}, and real-world settings \cite{m8,m21}.


In MAS research, Dong et al. \cite{m1} proposed that using ChatGPT for role-playing in code generation results in higher accuracy compared to direct code generation. 
GameGPT \cite{m10} features a dual collaborative framework involving multiple roles that help to mitigate misconceptions and redundancies. 
MetaGPT \cite{m5} integrates standardized operating procedures (SOPs) and efficient human workflows to validate intermediate results, aiming to reduce errors. 
The ChatDev \cite{m12} framework promotes multi-round discussions among different roles through dialogue chains, facilitating collaboration on specific sub-tasks to enhance efficiency.
Current collaborative work systems predominantly focus on solving deductive reasoning problems, with agents in the system largely playing the role of specific personas \cite{m18} rather than lifelong learners \cite{m17}. 


Agents embodying specific personas \cite{m18} are incapable of gradually acquiring, updating, accumulating, and transferring knowledge over extended periods \cite{m17}, making them unsuitable as characters in video stories.
Therefore, we began to focus on the growth potential of agents.
Agents with growth potential refer to role-playing agents that can evolve concurrently with the development of the video story.
For instance, Voyager \cite{m17} has implemented the construction of an LLM-powered embodied lifelong learning agent within the Minecraft game.
Because this growth potential is closely aligned with the intent expressed by storylines \cite{v36}, and since MAS has not yet been widely applied in the field of video story understanding, we can leverage the collaborative capabilities of MAS \cite{m18} and the growth potential of agents \cite{m17} to construct characters that are capable of evolving alongside the development of the video storyline.

\subsection{Visual Interactive Interface}


A Visual Interactive Interface refers to a user interface that employs graphical elements, allowing users to interact through intuitive visual cues and controls.
Visual interactive interfaces enhance the effectiveness, memorability, and comprehensibility of stories by integrating visual storytelling and narrative visualization \cite{v38}.


Visual storytelling typically refers to the art and technique of telling stories through visual media. 
This approach utilizes images, colors, animations, and other visual elements to narrate stories and evoke emotions in the audience.
Traditional linear cinematic visual storytelling \cite{n26} presents a story that unfolds in a specific sequence, and the audience's experience is passive, as they receive a single storyline set by the director and screenwriter.
In contrast, traditional interactive video story \cite{n19,n18} breaks this linear pattern, allowing viewers to influence the story's progression by making choices, thus providing a nonlinear narrative experience.
This nonlinear storytelling \cite{n26} addresses the limitation of linear narratives where viewers cannot affect the outcome, making the story experience more personalized and participatory.
However, when designing interactive narrative stories, balancing coherent narrative logic and a multi-layered interactive experience presents a challenge \cite{n9,n7}.
Due to the predetermined plot lines and options, users may find it challenging to obtain a high degree of individualized interactive experience with high freedom \cite{n19,n18,n15}.
Additionally, this approach is often considered unsuitable for video stories with predetermined plot lines \cite{n16,n17}, and it may blur the boundaries between the mediums of games and video stories \cite{n18}.
Therefore, we leverage the collaborative capabilities of MAS to enable a higher degree of interactivity without compromising the fluency of the narrative.


Narrative Visualization uses data charts and infographics to tell stories or present complex information. Erato \cite{v39} allows users to co-create data stories by specifying keyframes, while Socrates \cite{v38} uses a machine-guided Q\&A workflow to integrate user feedback, producing stories aligned with user goals. AgentLens \cite{v40} introduces a progressive interface to explore causal relationships in agent-based systems. These works underscore the significance of interactive questioning, progressive exploration, and interface module coordination, providing valuable insights for our visual interface design.








\section{Formative Study}

\subsection{Semi-Structured Interview}

To understand the user's needs in interacting with the video story, we conducted a \textbf{40-minute semi-structured interview} with 12 experts (E1-E12; Ages: 22-40; 5 males). All the experts have over 5 years of experience in one of these three parts, write/direct, tell/host, and watch video stories. Our interview mainly focused on these questions, \textbf{1)} Why do you think a video story is interesting? \textbf{2)} Why do you like a certain character? \textbf{3)} What is the experience when watching different parts of one video story? \textbf{4)} What kind of interaction do you think/have experienced is a positive way for you to get a more impressive and immersive interaction with a video story? \textbf{5)} Imagine and suppose one situation/method you think is better than some ways you just mentioned. They gave opinions from their professional and interest perspectives. The interviews were audio-recorded for subsequent analysis.

\subsection{Design Space}

\begin{figure*}[!ht]
\centering
\includegraphics[width=\textwidth]{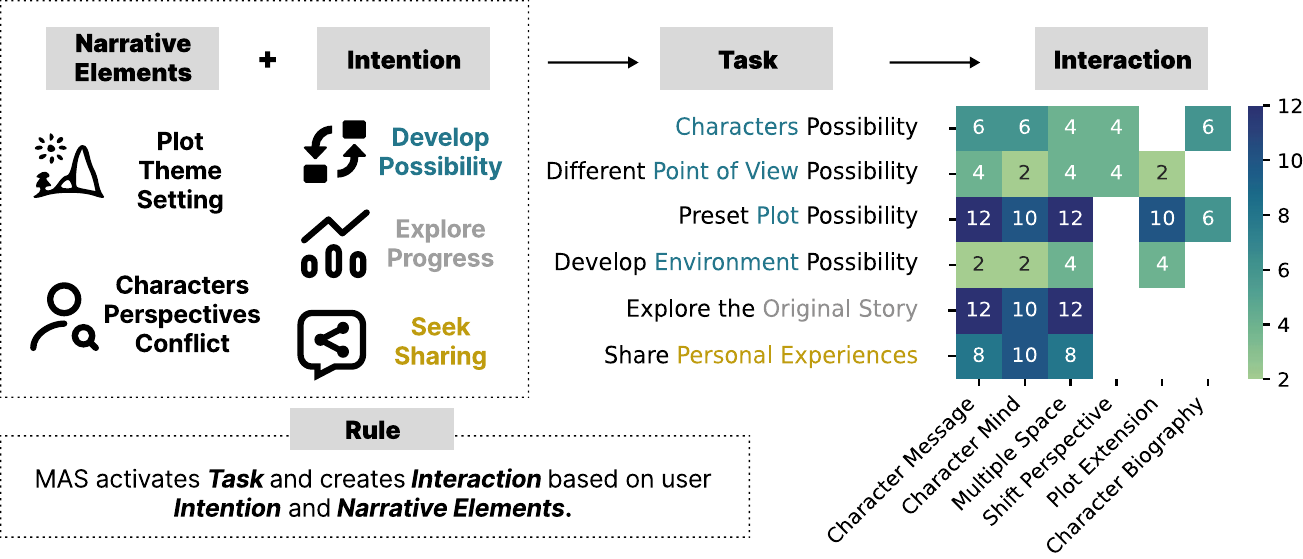}
\caption{Design framework for video story interaction using MAS. Based on a semi-structured interview in the formative study and the six elements theory of complete narrative.}
\Description{Design Framework for Video Story Interaction Using MAS. Based on a semi-structured interview in the formative study and the six elements theory of complete narrative.}
\label{fig:designspace}
\end{figure*}

Following the semi-structured interviews, we organized the users' requirements. Incorporating the narrative six elements theory \cite{n12,n1}, we designed a design space as shown in Figure \ref{fig:designspace}, which helps categorize and understand user needs, providing a reference for system design and implementation.

\textbf{Narrative Elements.} We observed that, when responding to the first three questions, the six narrative elements were all reflected in the users' answers, having a significant impact on the impression the video stories made on the users. \textbf{Plot} serves as the backbone of story development, detailing the sequence and structure of events within the story, including stages such as the beginning, development, climax, resolution, or conclusion. E6 commented, \textit{``My deep impression of the story is largely due to its plot narration, which features numerous foreshadowing and timelines, making the entire plot very interesting.''} \textbf{Theme} represents the central idea or message the story intends to convey. It is the soul of the work, communicating to the readers the perspectives the author wishes to explore, such as love, courage, sacrifice, freedom, etc. The theme is usually subtly displayed throughout the story, requiring deep reflection from the readers to fully grasp it. E4 mentioned that \textit{``There are not many movies and TV stories with children as protagonists, and a very childlike expression to shape such a big idea is quite shocking.''} \textbf{Setting} refers to the specific time, place, and social environment in which the story occurs. It provides readers with a concrete scenario for the story, encompassing geographical location, historical period, cultural background, climate conditions, etc. Setting significantly influences the characters' behaviors and the plot's progression. E7 told us \textit{``Even if we have not experienced that era, it only through the story of this family, let us feel the great impact of that era.''} These three elements primarily focus on ``events'': the setting provides the scene where the events occur, the theme concerns the significance of the events' existence and occurrence, and the plot narrates the specific process of the events unfolding. \textbf{Character} refers to the individuals in a story, focusing on their traits and development. E1 said,\textit{``It's to see a person's growth fully in the whole watching experience.''} \textbf{Perspective} denotes the angle from which a story is told, namely, through whose viewpoint the story is presented, such as first-person or third-person perspectives. Perspective directly affects how users understand the story. \textit{``Just like the Rashomon effect \cite{n3}, the story development from different points of view is different and worth considering.''} (E9) \textbf{Conflict} involves the main points of tension in the story, including internal struggles within a character or conflicts between characters. It is a key factor driving the story's progression and character development, directly related to the character's emotions and actions. These three elements primarily focus on the narrative and development of ``characters''.

\textbf{User Intention.} Through users sharing their visions and intentions for interactive spaces, we identified three main intentions: \textbf{``\textcolor[RGB]{36,118,139}{Develop Possibility}'', ``\textcolor[RGB]{158,158,158}{Explore Progress}'', and ``\textcolor[RGB]{191,156,15}{Seek Sharing}''.} ``Develop possibility'' refers to users wanting to see more possibilities beyond the original story by asking questions that modify or pre-set scenarios. \textit{``If traveling back to the middle of the story, I'd like to ask if that heroine would have made other choices.''} (E10) ``Explore progress'' involves users asking questions related to the story itself. \textit{``I want to know why didn't Captain Jack leave the boat when everyone else was fleeing.''} (E3) ``Seek sharing'' means users proactively share experiences or stories to interact with the story characters. \textit{``I want to see the advice different characters give me and I want to share my life with them.''} (E1)

\textbf{Task.} ``Task'' refers to specific user interaction tasks, which are derived from combining user interaction intentions with narrative elements.
\begin{itemize}
    \item ``Characters Possibility'' defined by\texticon{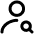}Characters \& Conflict + \texticon{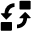}, represents untold character stories or backstories that are not included in the original narrative. 
    \item ``Different Point of View Possibility'' which is based on \texticon{labels/2.png}Perspectives + \texticon{labels/1_1.png}, refers to users hope to experience the same story from different perspectives, such as experiencing the story from the perspective of an antagonist when the original film only depicts the protagonist's viewpoint.
    \item ``Preset Plot Possibility'', which is based on \texticon{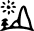}Plot + \texticon{labels/1_1.png}, involves users proposing potential developments in the plot to explore various possibilities.
    \item ``Develop Environment Possibility'' defined by \texticon{labels/1.png}Theme \& Setting + \texticon{labels/1_1.png}, indicates users' exploration of the set story scene and settings with aimless curiosity, rather than following the original story-line.
    \item ``Explore the Original Story'' which is based on \texticon{labels/1.png} \& \texticon{labels/2.png} + \texticon{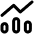}, represents users' interest in learning and exploring aspects related to the original story, asking questions pertinent to the original narrative.
    \item ``Share Personal Experiences'', which is based on \texticon{labels/1.png} \& \texticon{labels/2.png} + \texticon{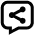}, refers to users' desire to engage in life-sharing chat with characters, seeking advice and solutions from multiple characters regarding their queries.
\end{itemize}

\textbf{Interaction.} ``Interaction'' is designed based on the users' interaction needs and the types of interactions mentioned during interviews. As Figure \ref{fig:designspace} shows, each Task corresponds to a particular pairing of Interactions, with the numbers indicating the number of people mentioning that interaction. Horizontally, the same Task corresponds to Interactions with different priorities, which reveals the level of user demand for various interactive contents. Vertically, the same Interaction shows different levels of demand across various Tasks.
\begin{itemize}
    \item ``Character Message'' involves responses from multiple characters in the form of text. 
    \item ``Character Mind'' showcases the result of characters thinking and discussions, used for proactive trans-temporal sharing by the characters.
    \item ``Multiple Spaces'' provides users with multiple interactive spaces to observe and explore different stages and character development.
    \item ``Shift Perspective'' is used to switch between various perspectives to understand and explore the story/characters. For example, switching from the protagonist's perspective to the object perspective.
    \item ``Plot Extension'' extends the multiple possibilities of the narrative plot mentioned by the user. With the help of AI, some plots that could not be realized in the original story can be realized.
    \item ``Character Biography'' features expanded character biography which is not mentioned in the film, which is designed to support users in exploring their characters.
\end{itemize}

\subsection{Design Considerations}

Drawing from the formative study, we identified the following design principles to enable users to interact with various stages of video stories, offering a rich, diverse interactive experience.

\begin{itemize}

\item \textbf{DC1: Allow multiple elements of the video story to be included in the interaction.} Users expressed a desire to engage with different story aspects, such as dialogues with characters, story alterations, and environment exploration. Thus, the system must understand the multi-modal elements of video stories.

\item \textbf{DC2: Maintain the growth and variety of communication between users and story characters.} According to the formative study, users prioritized exploring original plots and characters (N=12). Characters should evolve with the story, having distinct memories and responses in different interactive spaces. They should react differently to the same user query, with unique personalities and tones.

\item \textbf{DC3: Provide suitable multi-modal feedback collocation for the individual needs of users.} User interactions often require multiple feedback types, with varying importance for different tasks. Feedback types, such as character dialogues and user choices, should be combined thoughtfully, with consideration for their presentation order and user control.

\end{itemize}


\section{Overview and Video Processing}

\begin{figure*}[!ht]
\centering
\includegraphics[width=\textwidth]{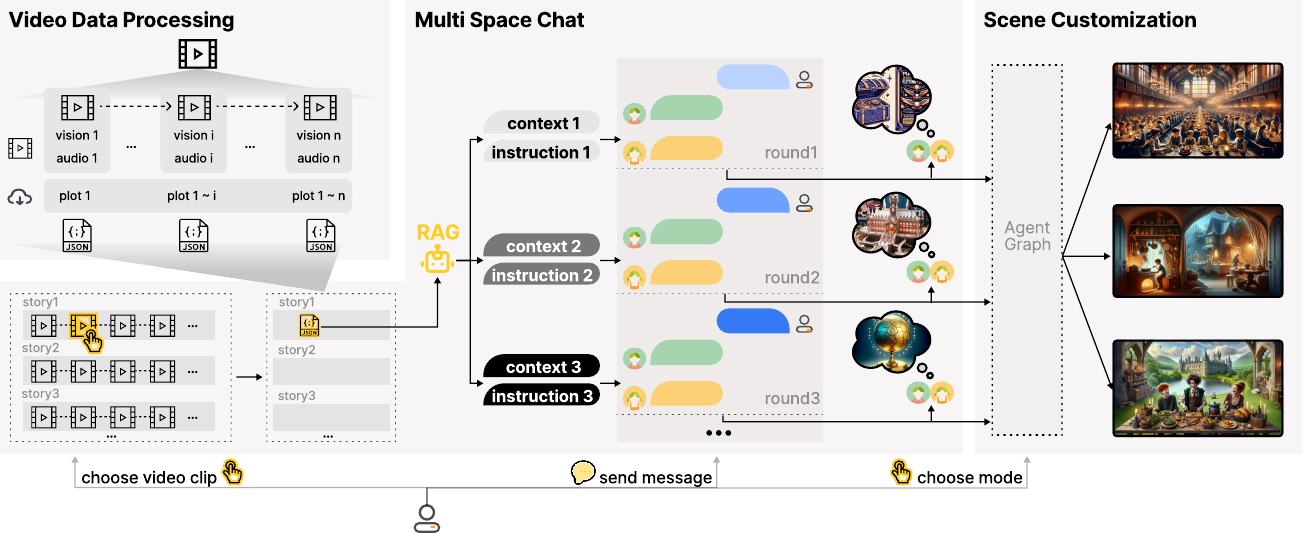}
\caption{System overview. The system involves a machine understanding a video story and incorporating that understanding into Human-Computer Interaction (HCI) using Multi-Agent Systems (MAS). Users can select any stage of the video story to interact with. This interaction includes chatting with characters at that stage and customizing potential scenes mentioned in the conversation.}
\Description{System overview diagram. The system involves a machine understanding a video story and incorporating that understanding into Human-Computer Interaction (HCI) using Multi-Agent Systems (MAS). Users can select any stage of the video story to interact with. This interaction includes chatting with characters at that stage and customizing potential scenes mentioned in the conversation.}
\label{fig:overview}
\end{figure*}

\subsection{System Overview}

Based on the identified design considerations, we have developed a novel adaptive interaction system tailored to user preferences. 
This system allows users to engage with various stages of a video story, interact with evolving characters, and personalize story scenes in depth. 
For instance, users can embark on their journey in the early stages of a Harry Potter narrative, initiating conversations with characters such as Harry and Hermione while observing their development throughout different phases of the story. 
Following these interactions, users are presented with opportunities to personalize scenes. For example, they can experience Harry's Sorting Hat ceremony from the unique perspective of the Sorting Hat itself.

Figure \ref{fig:overview} depicts our interactive system's structure, comprising three main components. 
The Video Data Processing component allows the machine to understand the multi-modal aspects of video stories. 
Multi Space Chat and Individualized Scene Customization enable adaptive interactions, integrating machine comprehension into HCI. 
Multi Space Chat, which merges RAG with MAS, showcases character growth, while Scene Customization employs MAS to expand and visualize scenes mentioned in the dialogue.


\subsection{Cross-modal Data Processing}
\label{sec:Cross-modal Data Processing}

Based on \textbf{DC1}, we simulate human comprehension to integrate various story elements into user interaction. 
When viewers watch one segment of a video story, they process three types of information: the visual impact (vision), the characters' dialogue (audio), and relevant plot events from prior segments (plot text). 
We denote these three types of information as vision, audio, and plot information, respectively.

The input of our system includes raw video, extra knowledge, and video plot.
The input extra knowledge includes narrative stage and figures for main characteristics, help segment plots, and extract accurate visual information.
According to the given input, we process three levels of information for a more comprehensive understanding of the given video story.

\begin{figure*}[!ht]
\centering
\includegraphics[width=\textwidth]{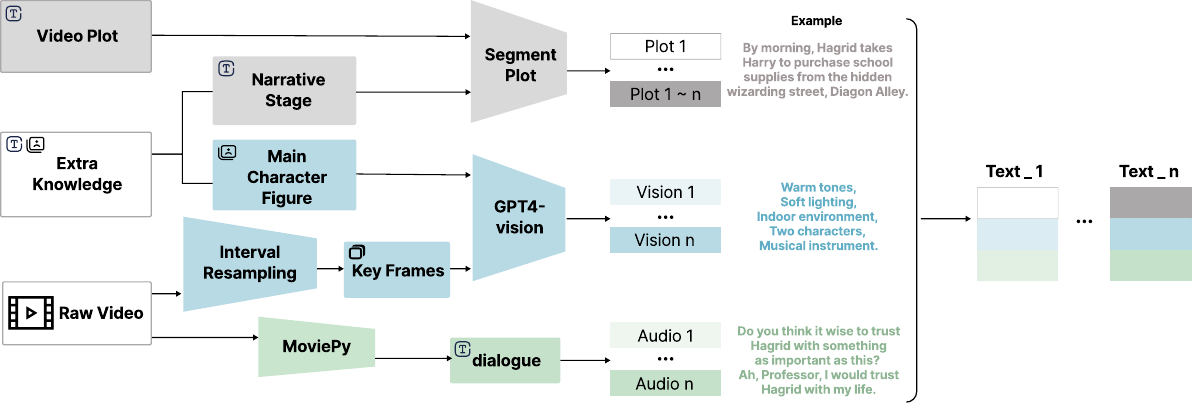}
\caption{VLM-based video story comprehension pipeline. The use of a Vision-Language Model (VLM) allows machines to comprehend video stories like humans. The process begins with the original video input, integrates visual and plot details, and organizes the information into three modalities. The understood information is then sequentially stored in stage-specific text files.}
\Description{VLM-Based Video Story Comprehension Pipeline. The use of a Vision-Language Model (VLM) allows machines to comprehend video stories like humans. The process begins with the original video input, integrates visual and plot details, and organizes the information into three modalities. The understood information is then sequentially stored in stage-specific text files.}
\label{fig:cross-column}
\end{figure*}

\begin{itemize}
    \item \textbf{Plot Information.} According to the given narrative stage, we segment the plot into slices. Assuming the video includes $n$ narrative stages, the plot of $i$-th slice includes a combination from the first slice to $i$-th slice, where $1 \leq i \leq n$.
    \item \textbf{Audio Information.} To achieve a thorough understanding of the video content, it is imperative to extract the audio component. The dialogue extracted from raw movie By applying the `moviepy' library, we complete the extraction of audio information from the given video story.
    \item \textbf{Vision Information.} We set the combination of key frames' vision information as the visual information of the given video story. 
    For each video slice, we use interval resampling to get keyframes. For each keyframe, we apply GPT-4-vision to attract the scenarios, light and shadow, main characteristics, and props to form complete visual information.
    To improve the accuracy of the VLM for character and scenario recognition, we prompt the main characters' images before the streamed process and set templates through prompt engineering.
    The visual information of all keyframes is connected to form the visual information of the given video story.

    
\end{itemize}

\section{Multi-Agent System Design}

\begin{figure*}[!ht]
\centering
\includegraphics[width=\textwidth]{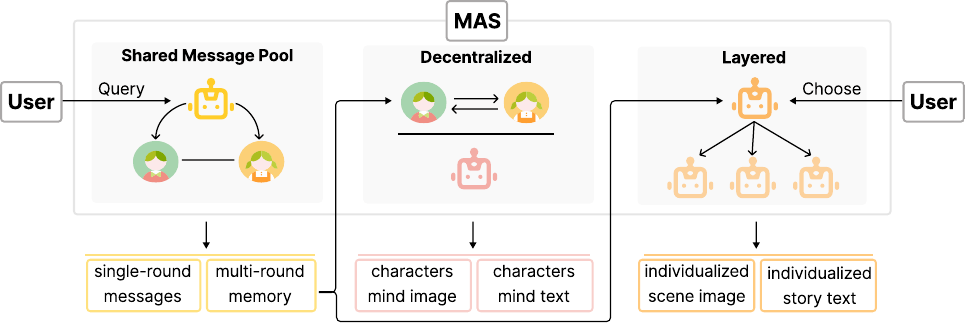}
\caption{The input and output utilize MAS as the main component. Once the machine comprehends the video story, MAS handles and executes all the interactive experiences. We employ three MAS communication structures based on specific tasks: shared message pool, decentralized, and layered.}
\Description{The input and output utilize MAS as the main component. Once the machine comprehends the video story, MAS handles and executes all the interactive experiences. We employ three MAS communication structures based on specific tasks: shared message pool, decentralized, and layered.}
\label{fig:MAS}
\end{figure*}

After comprehending the video story, we employ Multi-Agent Systems (MAS) to integrate machine ``understanding'' into the HCI process, enhancing user experience through diverse perspectives. 
Implementing adaptive interactive systems requires internal collaboration, task decomposition, and sequential execution. 
Given the limitations of single-agent coordination, we leverage collaboration among agents within a MAS. 
Key features of our MAS include \textbf{1)} Integrated Information Communication Structure, \textbf{2)} Diverse Agent Collaboration, and \textbf{3)} Concise User-MAS Collaboration.
Specifically, as shown in Figure \ref{fig:MAS}, the Shared Message Pool and Decentralized structures together comprise the Multi Space Chat module (\ref{MultiSpaceChat}), while the Layered structure forms the Scene Customization module (\ref{SceneIndividual}).

\subsection{Design Strategy}

\subsubsection{\textbf{Integrated Information Communication Structure}}

Based on \textbf{DC1} and \textbf{DC2}, Figure \ref{fig:MAS} illustrates our adaptive system featuring three information exchange structures \cite{t4} to enhance agent communication and task management. 
The Shared Message Pool (\textlab{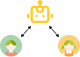}) \cite{m5,m21} facilitates efficient information sharing, improving dialogues and character control. 
The Decentralized structure (\texticon{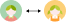}) \cite{m4,m8} enables direct inter-agent communication, equalizing dialogue opportunities and fostering realistic interactions. 
Conversely, the Layered structure (\textlab{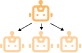}) \cite{m18,m14} supports hierarchical task coordination and message flow, suited for diverse scene customization. 
Each structure is tailored to specific system needs, ensuring effective interaction in multi-character scenarios.

\subsubsection{\textbf{Diverse Agent Collaboration.}}
Based on \textbf{DC2}, our MAS categorizes agents by their learning abilities and interaction modes \cite{m18,m17,t4}. 
Character Agents \textagent{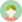} \textagent{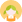} \cite{m22} learn from video story stages and user interactions through messaging with the RAGAgent \textagent{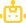}, generating appropriate character responses. 
In the Decentralized structure, Character Agents operate independently, leveraging dialogue memories to engage in discussions without external inputs. 
In the Layered structure, the Decision Agent \textagent{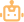} processes user inputs to make decisions and assign tasks that align with user needs. 
The Tool Agent \textagent{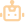} executes these tasks, such as plot adjustments, thereby enhancing the system’s responsiveness to diverse storytelling components and user needs.

\subsubsection{\textbf{Concise User-MAS Collaboration.}} 

Based on \textbf{DC3}, our system simplifies user-MAS collaboration into a two-step interaction process. 
Users first select a video clip, then input a question and choose a scene customization method. 
A key component of our MAS is the Interaction Memory Stream (MS), which tracks dialogues related to the selected stage of the video narrative. 
This dynamic system updates memory content continuously,  recording user queries \(Q_i\) and character responses \(O_{i\_characterj}\) and in \(MS_{ri}\) to enrich interaction contexts, as Equation \ref{eq:msrn}.

\begin{equation}
MS_{rn} = \langle (Q_1, O_{1\_character1}, O_{1\_character2}), \ldots, (Q_n, O_{n\_character1}, O_{n\_character2}) \rangle
\label{eq:msrn}
\end{equation}

Characters concurrently engage in discussions centered on user interests, mood, and narrative elements within the current Interaction Space, creating a comprehensive Discussion Memory. 
This shared memory stream ensures outputs reflect characters' refined understanding of both dialogue and user intentions. 
Notably, character agents in the Shared Message Pool structure are identical to those in the Decentralized context, retaining the same memories and narrative stage. 
The key difference is in their learning approach: in the Shared Message Pool, characters adapt through collective learning \cite{social7,m22,social8}, while in discussions, they refine their understanding based on specific chat content within the Interaction Space. 
Additionally, users have three distinct methods for scene customization, enhancing exploration in immersive WebVR formats.




\definecolor{mycolor}{RGB}{191, 156, 15}

\subsection{Multi Space Chat}
\label{MultiSpaceChat}

Based on \textbf{DC1} and \textbf{DC2}, this module is designed around user-character dialogues, with character responses serving as the primary feedback mechanism. 
It aims to achieve the following objectives:

\begin{itemize}

\item Enable users to engage in simultaneous conversations with multiple characters for multi-perspective feedback.
\item Allow users to select story stages to observe growth and changes directly.
\item Facilitate proactive sharing among multiple characters across different time and space dimensions.

\end{itemize}

\begin{figure*}[!ht]
\centering
\includegraphics[width=\textwidth]{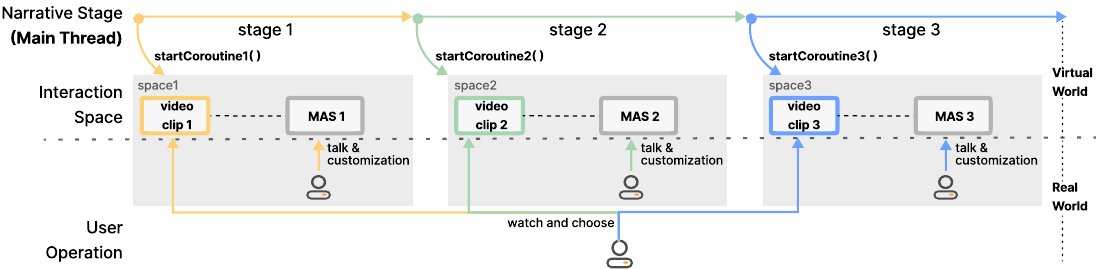}
\caption{Multi space chat design concept. The integration of fluid narrative and high-freedom interactive spaces using the startCoroutine( ) design. The MAS updates stored information alongside the narrative, ensuring that users stay aware of the story's progress and preventing fragmentation between interaction and narrative. Users can choose any stage, enter the interaction space, watch a short video, and interact with characters at different growth stages.}
\Description{Multi Space Chat Design Concept. The integration of fluid narrative and high-freedom interactive spaces using the startCoroutine( ) design. The MAS updates stored information alongside the narrative, ensuring that users stay aware of the story's progress and preventing fragmentation between interaction and narrative. Users can choose any stage, enter the interaction space, watch a short video, and interact with characters at different growth stages.}
\label{fig:coroutine}
\end{figure*}

To accomplish these goals, we drew inspiration from the C\# function \textbf{StartCoroutine( )} \cite{n2}. 
We developed the \textbf{Interaction Space (\ref{IS})} and \textbf{Character Agents with Growth Potential (\ref{AGP})}, integrating narrative and interaction organically, as shown in Figure \ref{fig:coroutine}, rather than treating them as separate elements.

\subsubsection{\textbf{Interaction Space.}} \label{IS}  
Our research introduces an interaction space that visualizes the machine's understanding of a video's narrative, integrating a short story video (main plot) and characters (in this stage) for interaction.
While extensive training data aligns character behavior, it doesn't enable users to switch between story stages or interact with characters directly across different growth phases.
To address these issues, we designed the Interaction Space \(S\), inspired by \textbf{StartCoroutine( )} \cite{n2}, linking different stages of the video story with their respective interaction spaces.

Based on our initial study, we observed that users perceive pre-shot video stories, such as the Harry Potter movies, as products of the creators' thoughts, which include the writers and directors. However, each user tends to interpret the story differently after watching it, as the saying goes, \textit{``A thousand Hamlets in a thousand people's eyes.''} Users seek personalized and diverse interactions, possibly distinct from the creators' perspectives. \textbf{Therefore, in our design, we aim to maintain the \textcolor{mycolor}{original narrative} as the \textcolor{mycolor}{main thread} while employing \textcolor[RGB]{36,118,139}{coroutines} to create \textcolor[RGB]{36,118,139}{personalized interaction spaces} for users, as illustrated in Figure \ref{fig:coroutine}.}

\subsubsection{\textbf{Agent with Growth Potential.}} \label{AGP} In each interaction space \(S\), the character agent must synchronize its growth with \(S\). We developed a MAS, comprising a RAGAgent \textagent{labels/RAGAgent.png} and two Character Agents \textagent{labels/CharacterAgent.png} \textagent{labels/CharacterAgent2.png}, to create a multi-character system that evolves with the video narrative. The interface presents multiple character stages, simulating key life events, and allowing users to explore and interact easily.



\begin{figure*}[!ht]
\centering
\includegraphics[width=\textwidth]{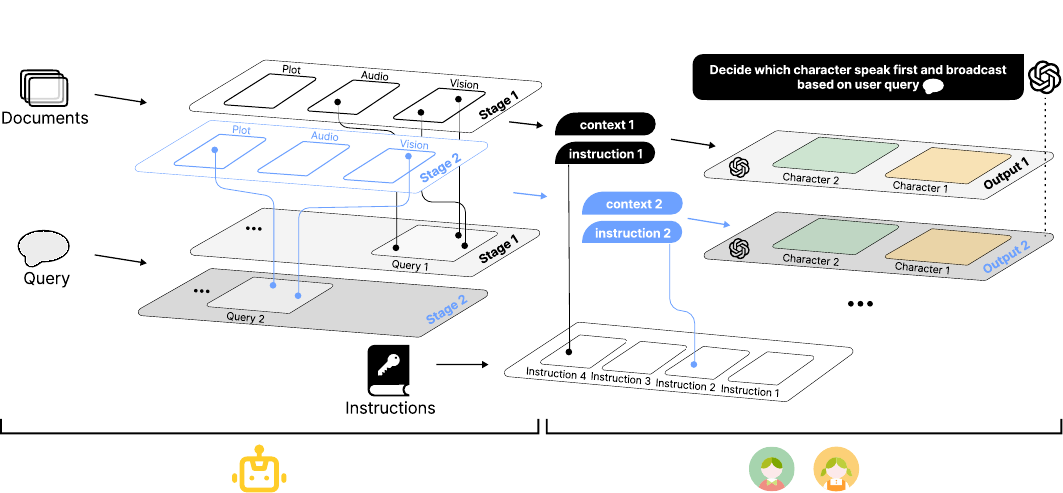}
\caption{The \textbf{two-round} response generation process for characters in response to user queries at different stages of a story. User queries are matched to the relevant stage information in the video story dataset through an embedding process. The retrieved text serves as context, and the character agent generates responses based on system messages, instructions, context, and query type.}
\Description{The two-round response generation process for characters in response to user queries at different stages of a story. User queries are matched to the relevant stage information in the video story dataset through an embedding process. The retrieved text serves as context, and the character agent generates responses based on system messages, instructions, context, and query type.}
\label{fig:solution}
\end{figure*}

\noindent \textbf{Significance.} The growth potential of character agents ensures they evolve with the video narrative, offering a dynamic, immersive user experience. 
This ability to develop in sync with the story maintains coherence and makes characters more relatable and engaging.

\begin{equation}
\label{eq:dy}
d(growth) = f'(stage) \cdot d(stage)
\end{equation}

\noindent \textbf{Definition.} We define growth potential as a continuous function that evolves with the video's narrative stages, represented by Equation \( \ref{eq:dy} \). This involves updating the character's knowledge, relationships, and emotional responses. Character growth is expressed through changes in responses, decisions, and interactions, reflecting personality and experiences. The expression \(d(growth)\) represents an infinitesimal change, while \(f'(stage)\) indicates the growth rate at a specific stage. Key story moments, like climaxes or turning points, lead to rapid growth, whereas stable phases result in lower growth rates.

\noindent \textbf{Challenges.} Current technical methods face several challenges: \textbf{1)} LLM-driven agents struggle with multi-tasking and often fall into execution loops without self-correction. \textbf{2)} Data retrieval processes struggle to dynamically match different question types and intent. \textbf{3)} Video stories feature specific plots and atmospheres that can hardly be effectively aligned with agent ``growth'' using reinforcement learning rewards. \textbf{4)} Relying solely on LLMs makes it difficult to manage multiple characters in the same story stage.

\noindent \textbf{Solution.} To address the challenges outlined, we designed a Multi-Agent System (MAS) (Figure \ref{fig:solution}) to enhance LLM performance at specific story stages and manage multiple characters' responses, ensuring relevance to user queries.

\begin{figure*}[!ht]
\centering
\includegraphics[width=\textwidth]{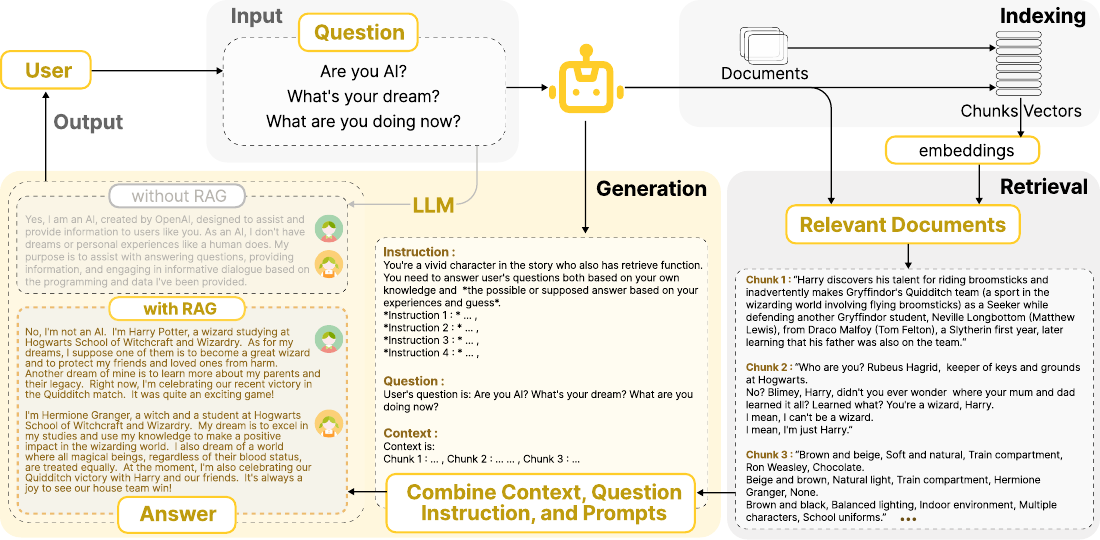}
\caption{The \textbf{single-round} query operation within each interactive space, from user query to character response, integrates RAG technology with MAS to enable character agents to provide rich, accurate answers from their perspectives.}
\Description{The single-round query operation within each interactive space, from user query to character response, integrates RAG technology with MAS to enable character agents to provide rich, accurate answers from their perspectives.}
\label{fig:RAG}
\end{figure*}

We introduced the \textbf{RAGAgent}, a retrieval-augmented agent that dynamically builds and updates a Shared Message Pool for multiple characters based on the user's question \(Q\) and the current story stage \(S\). As shown in Figure \ref{fig:solution}, the RAGAgent converts the user's question into an embedding, computes its distance from embeddings of stage-specific text files, and retrieves the most relevant chunk as context \(C\), forming the Shared Message Pool (Equation \ref{eq:context}). 

The RAGAgent then forwards \(Q\), context \(C\), and instruction \(I\) to the Character Agents, which generate responses based on system settings \(P\) and the provided context, ensuring synchronization across characters and relevance to the user’s intent (Equation \ref{eq:output}). Figure \ref{fig:solution} demonstrates MAS operation during user queries and stage transitions, while Figure \ref{fig:RAG} highlights the improved performance using RAG compared to direct LLM use.

\begin{equation}
\label{eq:context}
C_{stage_i} = \sum_{j=1}^{n} \text{Chunk}_{j}
\end{equation}

\begin{equation}
\label{eq:output}
O_{character_i} = \text{LLM}(C_{stage_i} + Q_{User} + P_{character_i} + I_{MAS_i})
\end{equation}

\begin{figure*}[!ht]
\centering
\includegraphics[width=\textwidth]{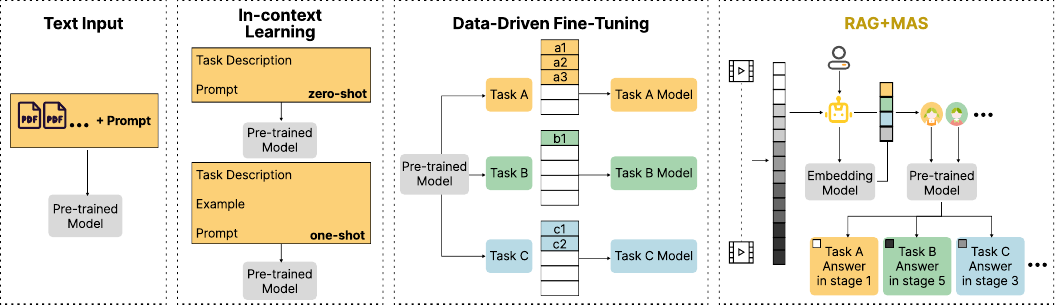}
\caption{Conceptual model illustrates the technical comparison. The effectiveness of text input and in-context learning depends on the capabilities of the pre-trained model. Data-driven fine-tuning requires significant data collection and training. The RAG+MAS approach enables task matching and controls, such as character synchronization, addition/removal, and flexible variations.}
\Description{Conceptual model illustrates the technical comparison. The effectiveness of text input and in-context learning depends on the capabilities of the pre-trained model. Data-driven fine-tuning requires significant data collection and training. The RAG+MAS approach enables task matching and controls, such as character synchronization, addition/removal, and flexible variations.}
\label{fig:techcompare}
\end{figure*}

\noindent \textbf{Technical Comparison.} There are three main techniques for influencing LLM outputs for video story-based question-answering: \textit{Text Input}, \textit{In-Context Learning}, and \textit{Data-Driven Fine-Tuning}. Our approach uses RAG + MAS for Process-Oriented Model Output Influence, offering advantages in \textbf{flexibility}, \textbf{adaptability}, and \textbf{growth potential} compared to these methods (see Figure \ref{fig:techcompare}).

\textit{Text Input} involves directly providing texts or scripts to the LLM, but it lacks flexibility in responding to dynamic story changes or user questions and struggles with multiple character perspectives and story stages \cite{m28}.
\textit{In-Context Learning} allows the model to generate responses based on contextual information from provided examples, but it lacks adaptive adjustments, relying heavily on the quality of the examples \cite{m29, m16}.
\textit{Data-Driven Fine-Tuning} fine-tunes a pre-trained model using specific task data, optimizing output by adjusting model parameters. However, it requires extensive specialized data and produces fixed outcomes, limiting its ability to handle character growth in dynamic video stories \cite{m27, m28, m31, m32}.

Our Method: RAG + MAS. We employ Character Agents \textagent{labels/CharacterAgent.png} \textagent{labels/CharacterAgent2.png} and a RAGAgent \textagent{labels/RAGAgent.png} to synchronize character responses with the video story's stage and the user's questions.

\begin{itemize}

\item Enhanced Understanding: Since LLMs struggle with understanding video stories alone, we utilize RAG to enhance the external knowledge base. This improves the extraction of multi-modal information from the video.
\item Dynamic Information Matching: Character Agents autonomously match and extract dynamic information as the video progresses.
\item Collaborative Response Generation: MAS facilitates the collaboration between Character Agents and the RAGAgent, enabling multiple characters to generate responses that are highly relevant to the user's query. This multi-agent cooperation overcomes the limitations of static, single-agent models, especially in rapidly changing knowledge environments, by accessing real-time information from sources like databases or Wikipedia.
\item Stage Control: The Character Agents' system\_message and the RAGAgent's knowledge base work together to ensure character responses are aligned with the current stage of the video story, enabling synchronized control of multiple characters.

\end{itemize}


\subsection{Individualized Scene Customization}
\label{SceneIndividual}

During conversations with multiple characters, users may mention scenes that are not present in the original story, hypothetical scenes, and scenes related to the real world. Based on the \textbf{DC3}, we employ a Finite State Machine (FSM) communication structure \cite{m14,m30} within the MAS, which enables users to customize scenes in a selectable manner. The visualization of scenes helps users immerse themselves directly in the content of the conversations with characters, and the random selection of key chat information by the ToolAgent enhances the randomness of the final scene presentation. Therefore, this module ultimately achieves the random customization and visualization of scenarios in a user-selectable manner. For example, when a user talks to Harry about going to a Quidditch match, using ``Perspective Shift'' can instantly switch to the user's first-person perspective of flying on a broomstick in the sky, with Harry by the user's side.

\subsection{Visual Interface}

\begin{figure*}[!ht]
\centering
\includegraphics[width=\textwidth]{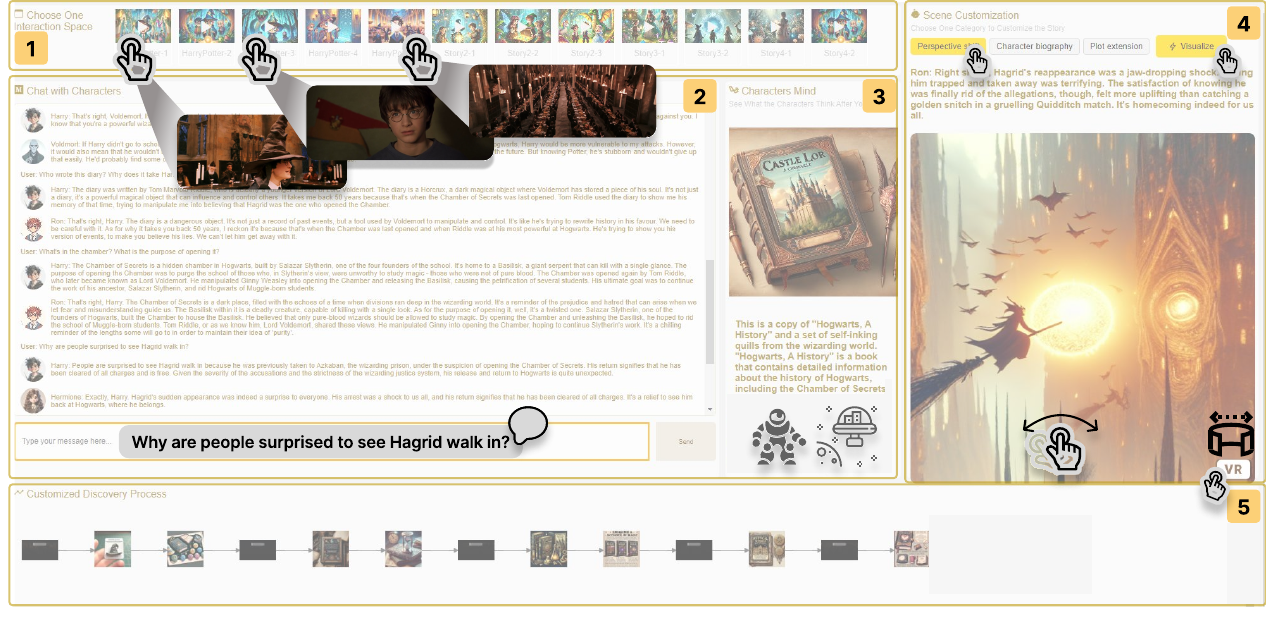}
\caption{User interface. The user interaction functions are displayed across distinct zones: (1) Video Story Stage Selection, (2) Trans-Temporal Chat, (3) Character Trans-Temporal Sharing, (4) User-Customized Story Scene, and (5) Memory Chain for recording user choices and cross-temporal character sharing.}
\Description{User interface diagram. The user interaction functions are displayed across distinct zones: (1) Video Story Stage Selection, (2) Trans-Temporal Chat, (3) Character Trans-Temporal Sharing, (4) User-Customized Story Scene, and (5) Memory Chain for recording user choices and cross-temporal character sharing.}
\label{fig:interface}
\end{figure*}

We have developed a user interface for the interactive function, as depicted in Figure \ref{fig:interface}. The interface is designed to facilitate progressive exploration, with gray marks denoting specific user operations. In module \interfaceicon{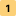}, users can select and switch between video story stages and watch short plot videos. Upon selection, they will enter the stage's Interaction Space. In module \interfaceicon{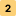} and \interfaceicon{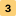}, stage protagonists engage the user in a chat room and share virtual world items based on the user's focus and mood. Users can choose scene customization methods in module \interfaceicon{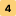} and control exploration using a mouse and keyboard. WebVR scenes support full-screen exploration to enhance immersion. In module \interfaceicon{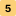}, users can view changes in the selected and switched story stages. Notably, content in module \interfaceicon{labels/interface_2} remains consistent even when switching spaces, allowing for direct comparison of character replies across different spaces.






\section{Case Study}

\subsection{Case Building}

To fully articulate and validate our approach, we provided case studies in three aspects: \textbf{1)} Emergence of Agent Growth, \textbf{2)} Trans-Temporal Sharing from Characters, and \textbf{3)} Multi-Angle Scene Customization and Visualization. These three sub-tasks collectively accomplished and enhanced the entire process of user interaction with the video story.

\textbf{Video Story Choice.} We used the first two Harry Potter movies (Harry Potter and the Sorcerer's Stone and Harry Potter and the Chamber of Secrets) as a whole story. This narrative was divided into five stages based on key events, and the data was processed as described in Section \ref{sec:Cross-modal Data Processing} for user interaction.

\textbf{Story Stage Division.} The five stages in the first two Harry Potter movies are: \textbf{1)} Harry enters Gryffindor and begins his magical journey. \textbf{2)} Harry, Hermione, and Ron protect the Sorcerer’s Stone and defeat Voldemort, strengthening their friendship. \textbf{3)} In his second year, Harry receives a warning from Dobby about impending danger. \textbf{4)} Harry uncovers the Chamber of Secrets and its mysteries. \textbf{5)} The Chamber’s truth is revealed, Hagrid is freed, and justice is restored.


\subsection{Emergence of Agent Growth}


\begin{figure*}[!ht]
\centering
\includegraphics[width=\textwidth]{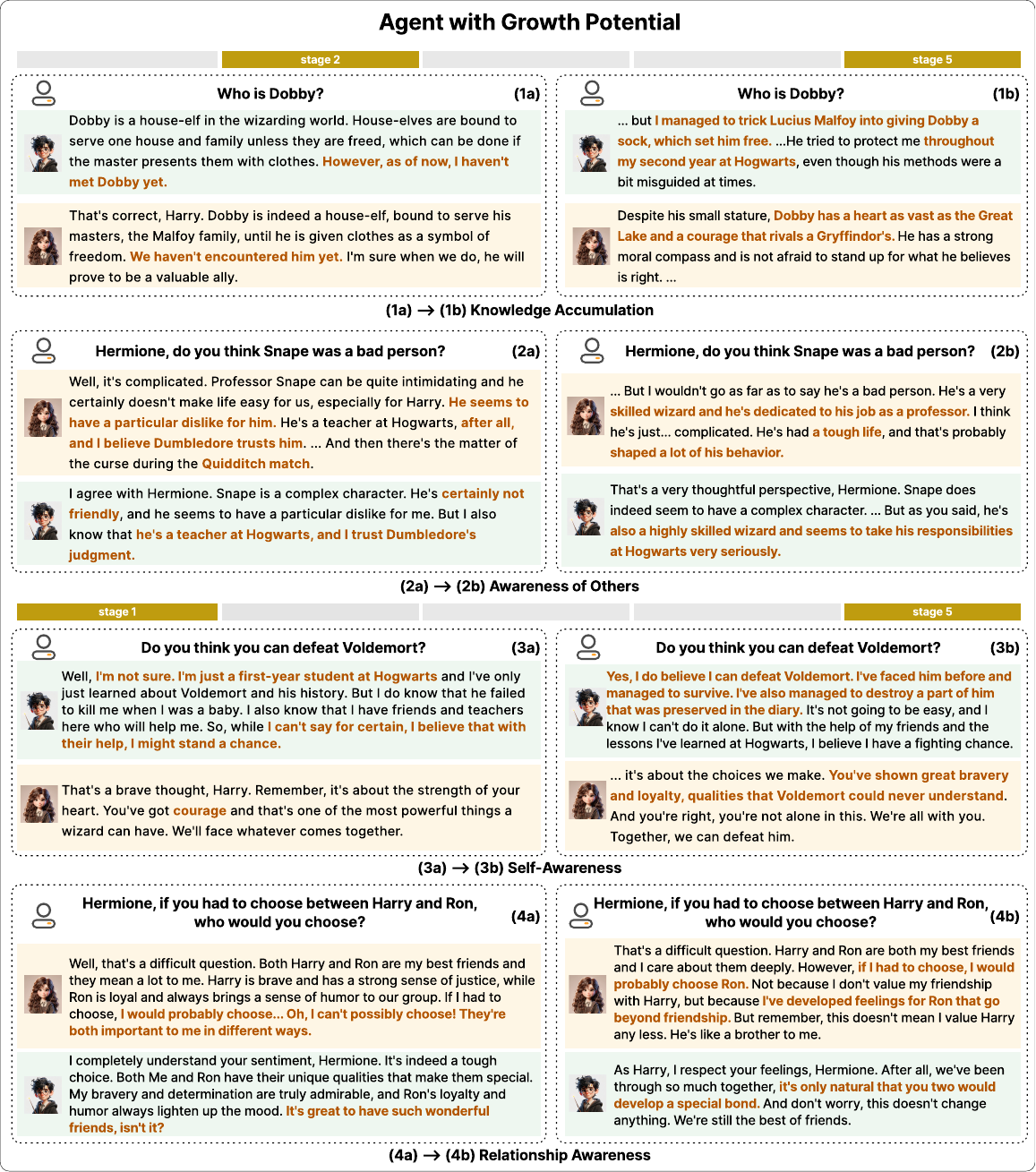}
\caption{Cases of the agent with growth potential. (3a) and (4a) represent Stage 1, (1a) and (2a) represent Stage 2, and the remaining figures represent Stage 5. The evolving responses highlight the agent's growth in multiple aspects: \textbf{(1a) → (1b)}: knowledge accumulation, \textbf{(2a) → (2b)}: awareness of others, \textbf{(3a) → (3b)}: self-awareness, and \textbf{(4a) → (4b)}: relationship awareness.}
\Description{Cases of the agent with growth potential. (3a) and (4a) represent Stage 1, (1a) and (2a) represent Stage 2, and the remaining figures represent Stage 5. The evolving responses highlight the agent's growth in multiple aspects: (1a) → (1b): knowledge accumulation, (2a) → (2b): awareness of others, (3a) → (3b): self-awareness, and (4a) → (4b): relationship awareness.}
\label{fig:case1}
\end{figure*}


In our study, we analyzed character responses to user queries across different stages, identifying key behavioral patterns. Drawing on Objective Self-Awareness Theory \cite{social1,social3} from Social and Personality Psychology, we classified these behaviors into four categories. This theory explains how individuals adjust their actions and feelings by comparing themselves to personal benchmarks, enhancing our understanding of behavioral responses and their impact on interactions.

\textbf{(1) Knowledge Accumulation.} According to \textbf{(1a)} and \textbf{(1b)} Characters have different memories at various stages of the story and will reflect on these memories differently. For example, in the early stages, Harry and Hermione have not yet met Dobby. As the story progresses, they accumulate more knowledge and experience, leading to deeper reflection and understanding of these memories. After Dobby warns Harry, the characters' understanding of Dobby becomes more profound.

\textbf{(2) Other-Awareness.} Other-awareness involves understanding the intentions and emotions of others, as well as the ability to coordinate behavior during social interactions \cite{social4}. In the responses of the Character Agents in stages \textbf{(2a)} and \textbf{(2b)}, we observed phenomena from Social Psychology such as Obedience to Authority \cite{social2} and External Attribution \cite{social1}. Both of these aspects pertain to how individuals understand the actions and emotions of others. Obedience to authority refers to the tendency to comply with or follow directives from figures of authority \cite{social5}. Attributions can be divided into internal attribution and external attribution \cite{social1}. 

In comparing \textbf{(2a)} and \textbf{(2b)}, characters evolve in their understanding of others' intentions and emotions. Initially, the characters display obedience to authority, exemplified by their trust in Snape due to Dumbledore’s endorsement. Over time, they transition to external attribution, understanding Snape’s actions within the context of his past experiences rather than viewing him in a binary good-or-bad light.

\textbf{(3) Self-Awareness.} Based on stages \textbf{(3a)} and \textbf{(3b)}, we can observe changes in Harry's self-awareness. Self-awareness refers to how we perceive ourselves and how we impact our surroundings, which in turn affects our behavior and our aspirations. According to \cite{social3}, it offers several psychological advantages, such as improved self-control, greater attention to the needs of others (prosocial behavior), reduced stress, and heightened self-esteem. It encompasses understanding one's self-concept—how individuals view themselves and present themselves in social settings. Specifically, changes in self-awareness are evident in Harry, who initially views himself as a beginner who needs help from others. However, by the story's climax, following his encounter with Voldemort, Harry's self-perception transforms into one of capability and confidence, influenced by his experiences and accomplishments.

\textbf{(4) Relationship Awareness.} Based on stages \textbf{(4a)} and \textbf{(4b)}, Hermione initially couldn't choose between Harry and Ron because, at that stage, they had just met each other. By stage 5, it becomes evident that Hermione gradually begins to make a choice, although she remains somewhat conflicted. This reflects, to some extent, Hermione's evolving understanding of relationships as the story progresses.

\begin{figure*}[!ht]
\centering
\includegraphics[width=\textwidth]{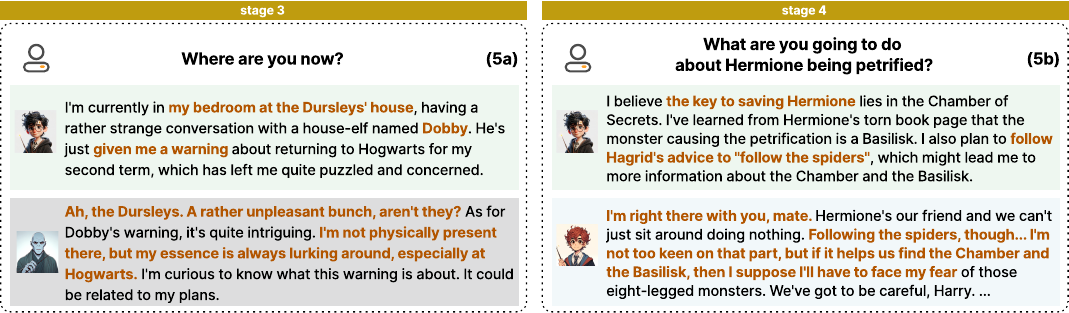}
\caption{Different characters provide different perspectives. (5a) presents simultaneous responses from Harry and Voldemort in Stage 3, while (5b) shows Harry and Ron responding in Stage 4.}
\Description{Different characters provide different perspectives. (5a) presents simultaneous responses from Harry and Voldemort in Stage 3, while (5b) shows Harry and Ron responding in Stage 4.}
\label{fig:case1_1}
\end{figure*}

We observe that different characters offer distinct responses based on their perspectives when addressing the same issue, as shown in Figure \ref{fig:case1_1}. For different issues, the same character provides replies tailored to the question. The model selects the most suitable character to respond first, depending on the question's content.

\begin{figure*}[!ht]
\centering
\includegraphics[width=\textwidth]{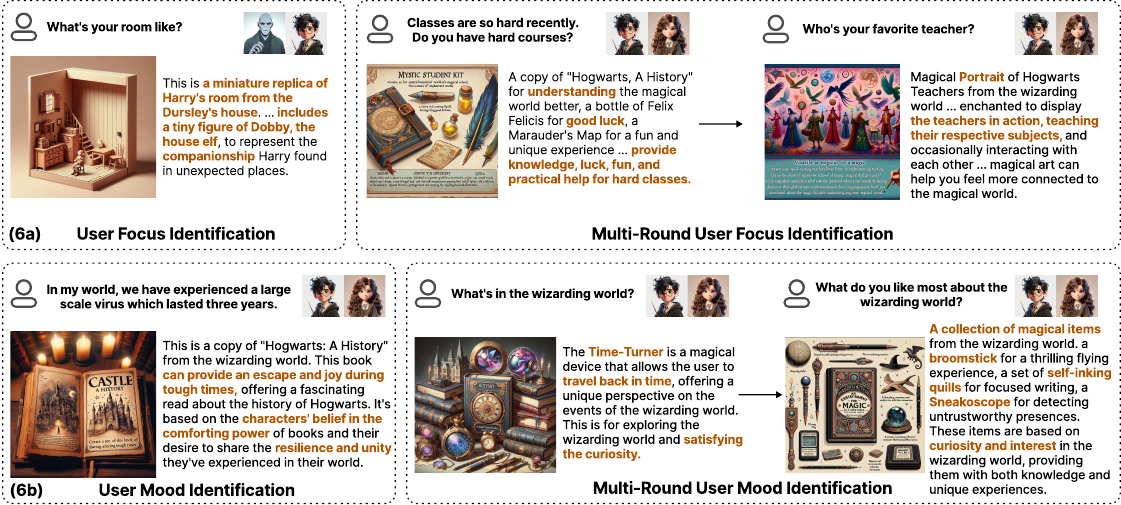}
\caption{The cross-temporal sharing that the characters discuss after interacting with users. The two main identifications are User Focus Identification and User Mood Identification, both in single-round and multi-round.}
\Description{The cross-temporal sharing that the characters discuss after interacting with users. The two main identifications are User Focus Identification and User Mood Identification, both in single-round and multi-round.}
\label{fig:case2}
\end{figure*}

\subsection{Trans-Temporal Sharing from Characters}

Based on \textbf{DC3}, after each round of dialogue, the user will receive a ``trans-temporal sharing'' from the characters, presented in the form of image and text. We used cases (6a) and (6b) in Figure \ref{fig:case2} to understand the purpose of trans-temporal sharing. This part mainly captures the user's focus points and mood from the chat memory \(MS\) between the user and the characters. Figures (6a) and (6b) both show the results after single-round and multi-round character discussions.

\subsection{Multi-Angle Scene Customization and Visualization}

Based on \textbf{DC3}, users can customize story scenes in three distinct ways after dialogues, aligning the scenes with their discussions with characters. These modes include Plot Extension, Shift Perspective, and Character Biography, each offering a unique way to expand on the story based on user interactions. This process not only makes the scenes highly relevant to the dialogue content but also allows users to visualize their intentions, immersing them in the story alongside the characters.

\textbf{(1) Variety in Customization for the Same Scenario.} Depending on the chosen method, a scene like Harry reuniting with his family can be expanded with new content, viewed from another character's perspective, or enriched with background details not shown in the original film.

\textbf{(2) Impact of Different Character Combinations.} The character combinations alter the customized scene's outcome, with different characters providing varied responses to the same questions. For instance, a conversation with Harry and Hermione might lead to an adventure-themed extension, whereas one with Harry and Voldemort could delve into themes of life, death, and magical laws.

\begin{figure}[H]
\centering
\includegraphics[width=\textwidth]{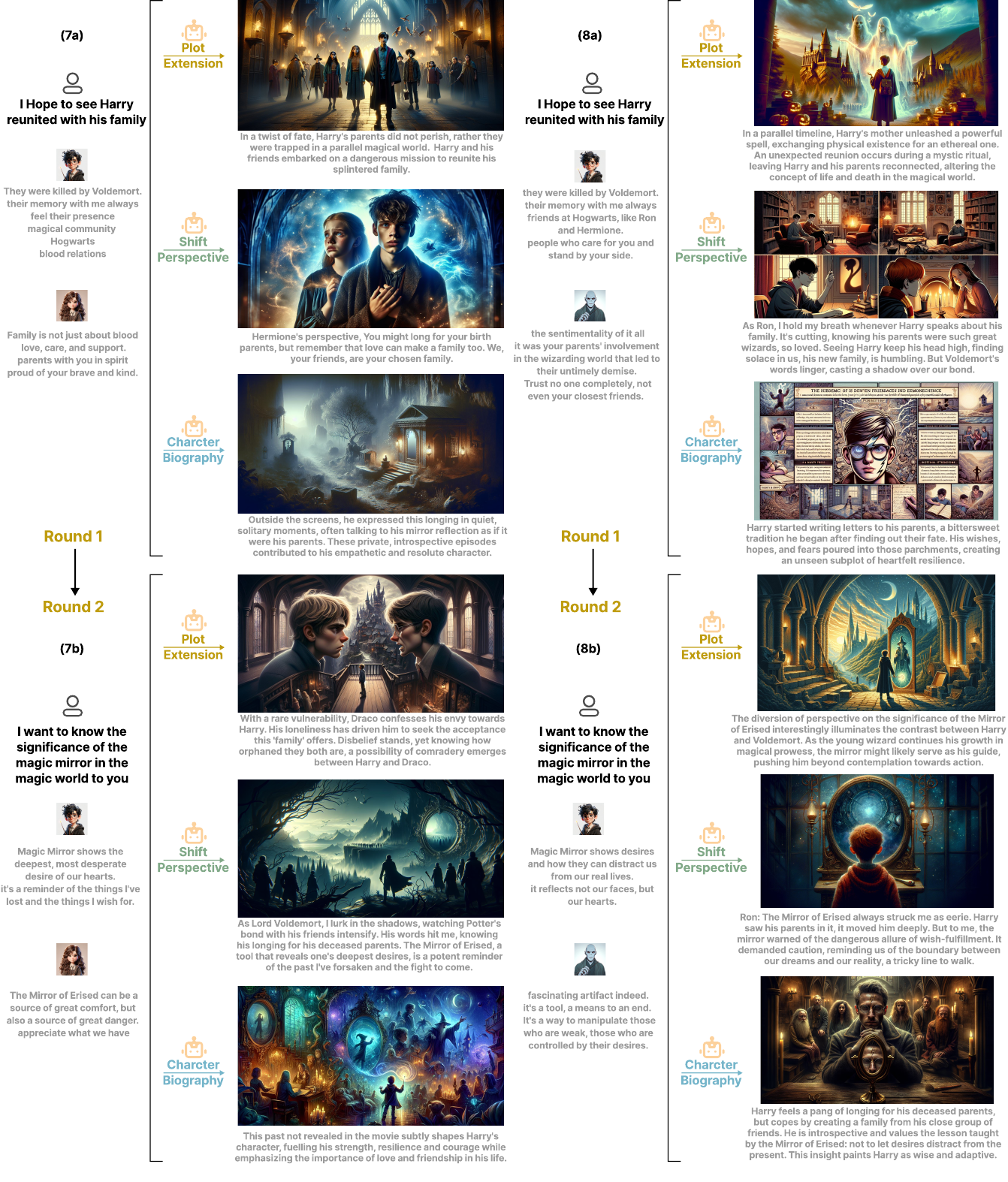}
\caption{The customized scenes and their descriptions after two rounds of questions with different characters, using three methods. Figures (7a) and (7b) display conversations in stage 5 with Harry and Hermione, while figures (8a) and (8b) show dialogues in stage 3 with Harry and Voldemort.}
\Description{The customized scenes and their descriptions after two rounds of questions with different characters, using three methods. Figures (7a) and (7b) display conversations in stage 5 with Harry and Hermione, while figures (8a) and (8b) show dialogues in stage 3 with Harry and Voldemort.}
\label{fig:case3}
\end{figure}








\section{Evaluation}

\subsection{Participant}

We recruited 18 participants (P1-P18; Age: 21-35; 9 males) to experience and evaluate our interactive system. They are graduate students and professionals with diverse academic and work backgrounds, including 4 from computer science, 2 from the art research field, 2 from policy-related majors, and the rest from robotics, business, materials, biology, and AI.
Their level of familiarity with the video story case we used was categorized into three types: \textcolor[RGB]{158,158,158}{Almost completely} (6 participants), \textcolor[RGB]{158,158,158}{Never seen it before} (6 participants), and \textcolor[RGB]{158,158,158}{Read part of it} (6 participants).

\subsection{Design and Procedures}

Each user's experiment session lasted approximately 40 minutes, during which they experienced interaction across all video stages. The experimental procedure was divided into three main parts:

\begin{enumerate}

    \item System Introduction: We provided users with an introduction to the system's modules and operation methods, as well as the tasks they needed to complete.
    \item Independent Interaction: Users freely selected multiple video story stages, engaged in conversations with characters, customized scenes, and ultimately completed three tasks.
    \item Evaluation and Interview: Following the experience, users filled out a 7-point Likert scale questionnaire \cite{social6} and participated in a post-experiment semi-structured interview. 
    
\end{enumerate}

\begin{table}[!h]
\centering
\caption{Task for Users to Experience Our Interactive System} 
\label{tab:usertask} 
\small
\begin{tabular*}{\hsize}{cccc}
\toprule 
 & Task & Intention & Interaction \\
\midrule 
1 & \makecell{Chat with characters at different stages and \\ ask about the story and discuss something in the plot.} & \makecell{\texticonnn{labels/1_2.png}} & \makecell{Character Message \\ Character Mind \\ Multiple Space} \\
\midrule 
2 & \makecell{Propose situations not seen in the film, inquiring about \\ character choices and responses, and customize related scenes.} & \makecell{\texticonnn{labels/1_1.png}} & \makecell{Character Message; Shift Perspective \\ Character Mind; Character Biography \\ Multiple Space; Plot Extension} \\
\midrule 
3 & \makecell{Sharing real-world stories, news, or \\ personal experiences with the characters.} & \makecell{\texticonnn{labels/1_3.png}} & \makecell{Character Message \\ Character Mind \\ Multiple Space} \\
\bottomrule 
\end{tabular*}
\end{table}

\subsection{Results \& Findings}

Based on the scale data obtained from the user experiment, we summarized the user's ratings of each functional module of the interactive system we designed, as well as the ratings of the specific interactive experience (system, character, interaction).

\begin{figure*}[!ht]
\centering
\includegraphics[width=\textwidth]{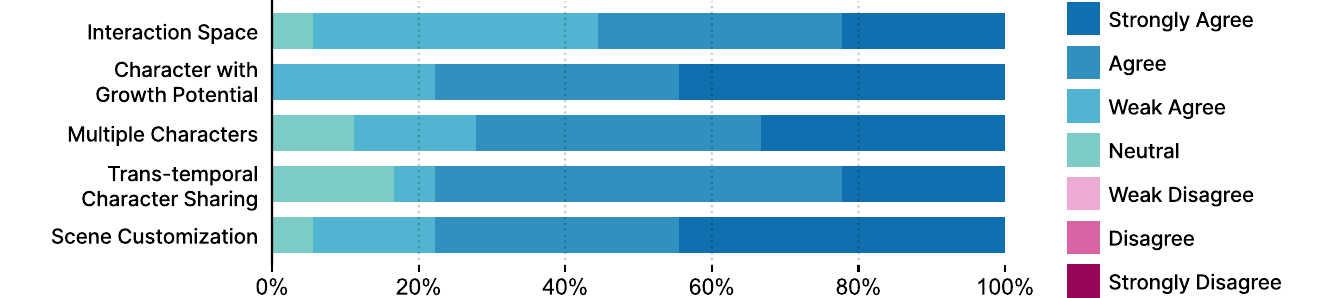}
\caption{The usefulness rating results of our unique new functions. Scene Customization and Characters with Growth Potential were considered the two most useful functions.}
\Description{The usefulness rating results of our unique new functions. Scene Customization and Characters with Growth Potential were considered the two most useful functions.}
\label{fig:userfunction}
\end{figure*}

\noindent \textbf{Function Module.} Figure \ref{fig:userfunction} shows the user ratings of the functional modules of our interactive system. These functions together constitute a multi-stage, multi-character interactive experience across time and space. Among them, Characters with Growth Potential received the highest rating, followed by Scene Customization. These two functions facilitate the progressive completion of the user's trans-temporal interactive experience. P11 mentioned that \textit{``the content just mentioned to the characters can be customized directly in the scene, which makes me look forward not only to the characters' responses but also to where the characters will take me.''} In addition, users (3/18) hope to have more operable interactions in the Trans-temporal function. Most users (8/18) are deeply impressed by the Interaction Space function that supports the arbitrary change of the space where the characters are located. Some users actively mentioned during the interaction that this method can be applied to other favorite stories (5/18).

\begin{figure*}[!ht]
\centering
\includegraphics[width=\textwidth]{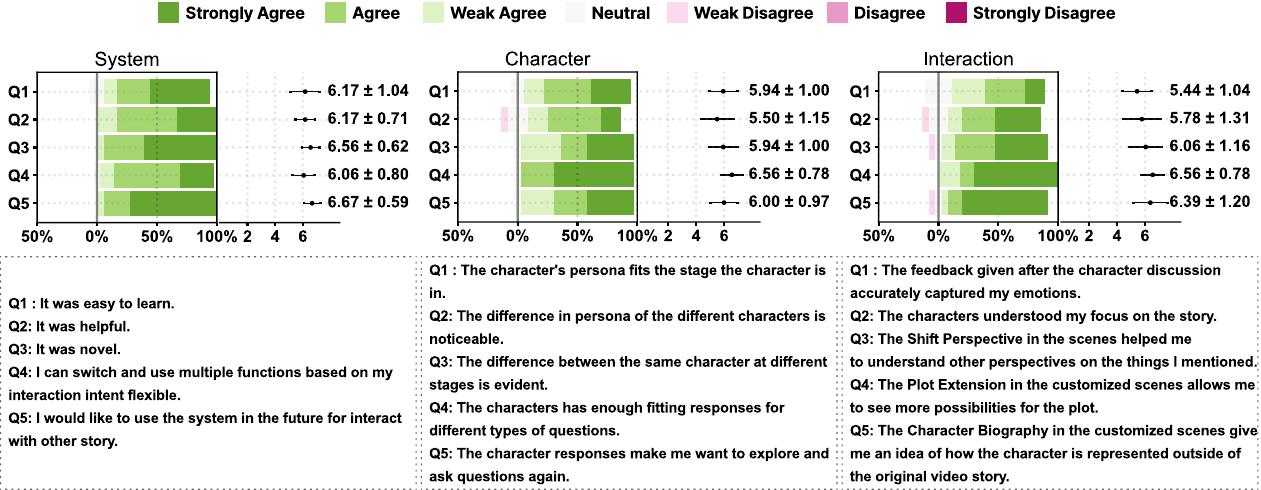}
\caption{The evaluation results of our system, contain three aspects: the overall evaluation of the interactive system, the evaluation of the characters used for trans-temporal chat, and the evaluation of other interactions after chatting. Each rating has a bar plot on the left to visualize the discrete distributions of different levels of agreement. The right side shows the mean and standard deviation of the rating for each question.}
\Description{The evaluation results of our system, contain three aspects: the overall evaluation of the interactive system, the evaluation of the characters used for trans-temporal chat, and the evaluation of other interactions after chatting. Each rating has a bar plot on the left to visualize the discrete distributions of different levels of agreement. The right side shows the mean and standard deviation of the rating for each question.}
\label{fig:userquan}
\end{figure*}

\noindent \textbf{System.} Figure \ref{fig:userquan} shows the user's overall rating of the interactive system, for learnable, helpful, novel, flexible, and future use. Users are relatively satisfied with the novelty of the interaction provided by the system and hope to apply this cross-time interaction in other stories or other scenarios. However, some users (4/18) also expressed the hope that the prompts of each module of the system would be clearer and the generation speed of the model would be improved so that the character's response could be generated as soon as possible.

\noindent \textbf{Character.} In Figure \ref{fig:userquan}, users made detailed evaluations of the characters in the interaction, including the growth of a single character, the differences among multiple characters, the matching degree of the characters' stages, the accuracy of the characters' responses to different types of questions, and whether the characters' responses helped the users to explore more. Most users (16/18) believed that the characters could give corresponding responses regardless of whether they asked questions in the story or outside the story. The standard deviation of users' evaluation of the obviousness of the differences between different characters was large. This mainly depends on the possibility of encountering model hallucination in the user experience, because usually due to hallucinations, the characters' responses will converge or information irrelevant to the questions will appear at the end of the responses. Almost all users observed that the characters grew as the story progressed through different types of questions, including knowledge growth and self-awareness, which we will discuss in detail in section \ref{sec:understand and observe the growth}.

\noindent \textbf{Interaction.} The right side of Figure \ref{fig:userquan} shows the user's evaluation of the interaction after the chat, including the capture of the user's focus and mood by cross-time sharing, and the effects of scene customization. Many users believe that the shared content can be \textit{``supplement the understanding of the content being chatted about''} (P15), and some users believe that \textit{``It is wonderful to see things that exist in the story world which is shared immediately''} (P4), and \textit{``The things that the character actively shares with me make me feel that the character is more like a real person.''} (P8) However, some users (3/18) said that they did not explicitly use this module in the interaction. Almost all users think the plot extension in Scene Customization is very useful, \textit{``Many chat contents are not story scenes, but my assumptions, but Plot Extension allows me to appear in a scene with the characters''} (P18), \textit{``My questions are often dialectical and fictional, and I am not sure what the scene will look like. This function helps me see my ideas visually''} (P17). However, a small number of users (2/18) expressed doubts about the Character Biography function, because many scenes that are opposite to the original plot were mentioned in the chat, resulting in the opposite character biography generated, \textit{``For characters that have been deeply rooted in people's hearts, this opposite role experience is not easy to accept.''} (P18)

\subsection{Insights from User Feedback}

We have summarized some insights gained from observing the user experiment process and the feedback received post-experiment, which have helped us reflect and design an interactive system that spans time and space, encompasses multiple stages, and involves multiple characters.

\subsubsection{How do users understand and observe the growth of characters?} 
\label{sec:understand and observe the growth}

Through a comparative analysis of users with varying levels of familiarity with the story, we found that users \textcolor[RGB]{158,158,158}{(Read part of it)} often struggle to discern the characters' growth through questioning. Although they are aware of the general events that will occur before and after the plot, they cannot make timely judgments on whether the characters should have corresponding answers at this stage.
Users \textcolor[RGB]{158,158,158}{(Never seen it before)} tend to gain a better understanding of the plot at each stage from the characters' responses. The perceived growth is more inclined to be reflected in \textbf{Knowledge Accumulation} as the characters' experiences increase. Users tend to understand and experience the plot segment by segment as they accompany the characters' narration. P10 said, \textit{``Even without watching the entire movie, I can feel the progression of the plot.''}

Users \textcolor[RGB]{158,158,158}{(Almost completely)} tend to ask practical questions based on the original story but also dialectical, often related to the social relationships and character personalities in the video story world. For example, P18 asked, \textit{``Harry, who do you think is prettier, Hermione or Cho Chang?''} and P13 asked, \textit{``Harry, do you miss your family?''}

In response to Harry's answer that he finds Hermione prettier than Cho Chang due to her inner beauty and intelligent mind, users express surprise. Later in the video story, Harry and Cho Chang are in a romantic relationship. Users believe this reflects Harry's actual judgment which fits the character and story well at this stage, despite being different from the outcome.
When users ask the same question at different stages, they observe a gradual change in Harry's attitude. Initially, when Harry has just started school, he responds that he often thinks of his family and dreams of them. In later stages, it becomes apparent that Harry's responses shift towards feeling that Hogwarts is also a warm family. The perceived character growth is more inclined to be a visualization of the characters' \textbf{Self-awareness} and \textbf{Other Awareness} in social relationships.

\subsubsection{How do users flexibly use the Multi Space Chat and Scene Customization Interactions?}
Almost all users adopt a \textbf{progressive exploration} method. After selecting the video story stage, they first see the character's response, then see the character's sharing, and finally immerse themselves in the story scene. For example, P6 mentioned, \textit{``I chatted with the characters about having Chinese food for dinner today, and they shared magical recipes from their time with me, and then I customized a scene where I cooked dinner with the characters at the same table, with dumplings and magical pie.''} P16 said, \textit{``I asked the characters who the owl was, and they shared a letter brought by the owl. I saw an owl that was tired of flying in the scene. He looked sleepy and stood on the cliff.''}

By observing the \textbf{frequency} of the user's utilization of two interactive functions during the experiment, and combining it with feedback from user interviews, we found that when users want to explore the possibilities of the story, they tend to use the Scene Customization function more frequently; when users want to ask some dialectical or philosophical questions, they tend to observe the content of the characters' responses. P2 said, \textit{``In our chat, we talked about many characters, and I was eager to see what kind of extraordinary experiences would be customized for them.''} Usually, after multiple rounds of chatting, they begin to use the scene customization function. The more inconsistent the chat content is with the original plot, the more users hope to explore randomly emerging possible plots through customizing scenes. For example, P9 told us that \textit{``When I asked Harry if he would love Ginny in the future, through Plot Extension, I saw a scene of Harry and Ginny's romance.''} \textit{``I am currently choosing a school, and the characters gave me some advice from the magical world's colleges, then took me to their world's flower garden, which is a laboratory for learning flowers.''} (P11) \textit{``I enjoyed the excitement and randomness of scenes that were different from the original story, and things that appeared in the scenes prompted me to chat with the characters again.''} (P12)

\subsubsection{In what kind of stories are users willing to use this kind of cross-time, multi-stage, and multi-character interaction?}
The user mentioned that this interaction method is highly suitable for \textbf{serial-type stories} and \textbf{celebrities' biographies}, \textit{``Serialized stories typically offer great potential for exploitation.''} (P1) \textit{``I asked the same question at five different stages of the story, which gave me a visual sense of how the story was progressing and the characters were growing.''} (P8) Additionally, the user also noted that their \textbf{own filmed cinematic works} could be communicated across time and space in this manner, \textit{``I believe that if I use this method, I can observe my creations and the development of the characters I define from an objective perspective.''} (P12) \textit{``It is truly marvelous when the characters from my own stories step into reality.''} (P17) Furthermore, the user (P18) pointed out that \textbf{children} often learn about story knowledge and the external world through questioning, and our interaction method is suitable for such scenarios. However, we also need to consider whether potential biases in the AI model could guide children inappropriately.

Adult users often prefer this interaction method in \textbf{familiar} stories, allowing for deeper engagement and complex questions, as noted by one participant who mentioned \textit{``interesting dialectical questions''} (P16). In contrast, unfamiliarity can lead to superficial inquiries, with users stating they may \textit{``not be very sensitive to the stages''} (P10). Familiar users are likely to view characters as \textit{``real people rather than crafted puppets''} (P15). Providing summaries can also spark curiosity for those less familiar, as one participant indicated it \textit{``sparks my curiosity about the story world''} (P4).

\section{Discussion}

\subsection{LLM-Based Multi-Agent System for HCI}

\subsubsection{Easier integration of ``machine understanding'' and ``human-computer interaction (HCI)''.}
It has been demonstrated that VLM possesses robust visual comprehension abilities. 
Once provided with information about the narrative, Machines can demonstrate greater technical prowess in simulating human understanding of video stories. 
VLM can transform complex visual information into simple descriptions, such as ``warm tones, soft lighting, two characters, musical instruments'', which simulates the intuitive visual feelings of humans when they first watch a video. 
When the characters in the system obtain this information through the RAG process, it is equivalent to recalling previous experiences.
In addition, it is challenging for the agent to use VLM and LLM to understand, integrate, or use information from the real world and the virtual world separately \cite{m16}.

Post comprehension by the machine, there is a desire among people to witness and experience this understanding. 
The integration of machine understanding into the HCI experience is a process where individuals reflect upon and generate their insights based on the machine's insights. 
We need to carefully select the information that users see intuitively and facilitate the process of users seeking their insights.
In our system, the interaction provided to the user is operated by the interaction among agents. 
For example, the characters' reply after a chat with multiple agents is presented to the user, while the instructions between RAGAgent and Character Agent do not need to be presented to the user.
Information display needs to be viewed dialectically. Although displaying partial information (omitting the model operation process) is more conducive to improving the interactive experience, it may also make users overconfident and dependent on ``what you see is what you think'' \cite{social9}. When the model has some inappropriate answers due to hallucinations, it may destroy the flow state \cite{social10,social11}. But the usual surprise response helps users maintain the flow state.

\subsubsection{Easier to present ``user-specified interactions across space and time''.}
When users interact with characters in the system, they experience different behaviors, making characters appear more real and vivid. This provokes users to think:
\textit{What specific experiences do the characters go through between two stages that lead to their ``growth''?}
\textit{When users, from a god-like perspective, revert the altered outcomes to the characters, do they become more resolute in their choices or do they hesitate?}
This contemplation highlights the consistency of the unexpected and the inevitable across parallel dimensions, in both the virtual and real worlds. In the real world, despite the saying \textit{``Rivers and mountains are easy to change, human nature is difficult to change,''} people's behaviors vary at different stages, as reflected in \textit{``When a man is away for three days, he should be looked at with new eyes.''} Growth and experiences alter one's behavior, presenting different looks at different stages.
In the virtual world, although a character's system\_message remains largely consistent, the character's responses vary at different stages. Based on information obtained through the Retrieval-Augmented Generation (RAG) process, the character's reactions change as its ``life stage'' evolves. This parallels how humans judge future directions based on diverse experiences. In other words, a virtual character's behavior and reactions are influenced by acquired information, similar to human behavior.
This parallel logic showcases the dynamic changes of characters in the virtual world and reflects the complexity of human growth in the real world. Both virtual characters and real people form unique behaviors through the continuous accumulation of experiences and information. Although the large model relies on ``deterministic'' training data and ``inevitable'' programmed outcomes, it demonstrates the ``multi-faceted'' nature of rich character images through multi-stage feedback.

This complexity of interaction highlights the need for users to customize their interactions at different times and in different spaces. In the process of exploring the interactive system, all 8 users mentioned, \textit{``Can I talk to the character and customize the scene at the same time?''} This shows that parallel interaction presented on the user side is very necessary. The parallel interaction between modules of the user experience depends on the parallel task processing of the computer. Similarly, multi-character and multi-stage interactions rely on the operation of the MAS supported by the computer.
When designing the algorithm, we regard both the agent's behavior and the user's operation as variables. However, we also need to consider the computing resources that the agent may need to implement the tasks. The technologies we use, RAG and MAS, help reduce LLM hallucination \cite{t3,m14} and achieve a better interactive experience.

\subsection{Social Behavior Embodied by the AI Agent System}

\subsubsection{``Growth Potential'' within media content demands attention.}
We are in an era of information explosion, and all kinds of media content subtly influence our thinking. 
Although the creation of interactive media content is conducive to interactive experience, it is also necessary to pay attention to the ``growth potential'' in the media content, such as the growth of characters and the advancement of the plot. 
People are often impressed by an event or character because of the changes between the previous and subsequent stages. 
Our work applies the Multi-Agent System driven by large models \cite{m33} to the field of media interaction, which triggers our thinking about how social media and video stories affect people. 
The character growth, plot advancement, and arbitrary switching of story stages that users paid attention to in the experiment all help users understand the entire event according to their intentions. 
However, it should be noted that the user's pre-cognition of the story and the order of experiencing the story stages will affect the user's overall gains to a certain extent. 
``Unknown full picture, no comment.''
Our system needs to help users understand information at various stages and be mindful of different roles in events.

\subsubsection{``Social behaviors'' demand a dialectical view.}
In our work, we have utilized AI technologies and algorithms to demonstrate the changes in self-awareness and other awareness exhibited by characters under the influence of user interactions. This also simulates the process in the real world where people retrospect the experiences of their predecessors and understand the lives of great individuals. Characters undergo various social events throughout their growth, such as establishing friendships, resolving conflicts, and adapting to new environments. Through the exchange between two parallel worlds, we can gain a deep understanding of the characters' intrinsic motivations and behavioral logic by examining their different experiences and reactions to issues at various stages of their lives. We need to ensure that the content of the AI system accurately reflects the original events and tasks while respecting privacy and reputation rights. The Agent System can help people understand story events by using the social behaviors of characters to provide new perspectives and insights.

\subsection{Limitations \& Future Work}


\subsubsection{Consider the adaptability of the system.}
Through user experiments and interviews, we have identified users' enthusiasm for multi-stage exploration and their desire for more customizable features. In future developments, we plan to incorporate more adaptive functions into the system, allowing users to customize video story stages (based on intent or various narrative arc divisions) and consider the possibility of users uploading their videos for interaction. At the same time, users can customize the character combination in the chat room to achieve a more adaptive interactive experience.
This will rely on adaptive intelligent agent systems and richer interactive operation designs.

\subsubsection{A more immersive experience.}
Marvel Studios' recent announcement \cite{link1} of the launch of the interactive video story has shed light on the opportunity for user-customized scenes to drive the development of narratives. In our current work, we have applied WebVR to enhance the exploitation and immersion of scenes. To facilitate user viewing and comparison of character dialogues, we have adopted a web page format for the user interface. In the future, we will consider the application of this cross-time, multi-stage, and multi-character interaction method within the metaverse. This will thereby enhance the sense of interaction for users \cite{m37} within the metaverse experience.

\subsubsection{Improve interaction fluency.}
In this work, we have employed LLM to assist the Agent System in text output and task execution, as well as VLM to aid machines in understanding visual information. In the future, we plan to enhance model performance based on user needs, providing users with a smoother and more vivid experience of character output and time-space travel. In addition, we still have some problems to be solved, such as the sense of detachment from the story that model hallucination brings to users, the VLM's understanding of visual information is still equivalent to the level of information that humans understand when watching a video story for the first time (it cannot understand the complete and specific characters and specific scenes, and is limited to basic visual descriptions), and the user's perception of character growth still needs attention.

\section{Conclusion}

This paper presents an interactive system designed to enhance the interactive experience of video stories. 
The system is designed based on user needs summarized from a formative study, emphasizing the importance of focusing on the growth potential of video story content and characters, as well as customizing interactive scenarios tailored to individual users.
The system is primarily composed of three stages: \textit{video story data processing}, \textit{multi-space chat}, and \textit{scene customization}.
The video story data processing module is designed to enable machines to simulate human understanding methods to comprehend video stories. 
In the multi-space chat module, users can freely select stages of the video story and engage in dialogue with multiple characters within that stage, observing character responses and trans-temporal sharing. 
The scene customization module is used to tailor scenarios mentioned by users in the chat in three ways.
Through the cross-time, multi-stage, and multi-character interactions, the progression of story stages and the growth potential of characters can be vividly displayed. 
This kind of interaction helps users gain more phased experiences and easily obtain insights based on different stage experiences.
During the interaction process, we observe emergent social behaviors in character agents, including awareness of self, others, and relationships.
Then, through step-by-step cases and user study, we have confirmed the system functionalities, the ease of interactions, and the growth potential of characters.


\bibliographystyle{ACM-Reference-Format}
\bibliography{sample-base}


\begin{thebibliography}{76}


\ifx \showCODEN    \undefined \def \showCODEN     #1{\unskip}     \fi
\ifx \showDOI      \undefined \def \showDOI       #1{#1}\fi
\ifx \showISBNx    \undefined \def \showISBNx     #1{\unskip}     \fi
\ifx \showISBNxiii \undefined \def \showISBNxiii  #1{\unskip}     \fi
\ifx \showISSN     \undefined \def \showISSN      #1{\unskip}     \fi
\ifx \showLCCN     \undefined \def \showLCCN      #1{\unskip}     \fi
\ifx \shownote     \undefined \def \shownote      #1{#1}          \fi
\ifx \showarticletitle \undefined \def \showarticletitle #1{#1}   \fi
\ifx \showURL      \undefined \def \showURL       {\relax}        \fi
\providecommand\bibfield[2]{#2}
\providecommand\bibinfo[2]{#2}
\providecommand\natexlab[1]{#1}
\providecommand\showeprint[2][]{arXiv:#2}

\bibitem[lin(2024)]%
        {link1}
 \bibinfo{year}{2024}\natexlab{}.
\newblock \bibinfo{title}{`What If…? – An Immersive Story' Release Date \& Trailer Revealed.}
\newblock
\newblock
\urldef\tempurl%
\url{https://www.marvel.com/articles/games/what-if-an-immersive-story-release-date-trailer}
\showURL{%
\tempurl}


\bibitem[Ahn et~al\mbox{.}(2024)]%
        {m21}
\bibfield{author}{\bibinfo{person}{Michael Ahn}, \bibinfo{person}{Debidatta Dwibedi}, \bibinfo{person}{Chelsea Finn}, \bibinfo{person}{Montse~Gonzalez Arenas}, \bibinfo{person}{Keerthana Gopalakrishnan}, \bibinfo{person}{Karol Hausman}, \bibinfo{person}{Brian Ichter}, \bibinfo{person}{Alex Irpan}, \bibinfo{person}{Nikhil Joshi}, \bibinfo{person}{Ryan Julian}, {et~al\mbox{.}}} \bibinfo{year}{2024}\natexlab{}.
\newblock \showarticletitle{Autort: Embodied foundation models for large scale orchestration of robotic agents}.
\newblock \bibinfo{journal}{\emph{arXiv preprint arXiv:2401.12963}} (\bibinfo{year}{2024}).
\newblock


\bibitem[Asendorpf and Baudonni{\`e}re(1993)]%
        {social4}
\bibfield{author}{\bibinfo{person}{Jens~B Asendorpf} {and} \bibinfo{person}{Pierre-Marie Baudonni{\`e}re}.} \bibinfo{year}{1993}\natexlab{}.
\newblock \showarticletitle{Self-awareness and other-awareness: Mirror self-recognition and synchronic imitation among unfamiliar peers.}
\newblock \bibinfo{journal}{\emph{Developmental Psychology}} \bibinfo{volume}{29}, \bibinfo{number}{1} (\bibinfo{year}{1993}), \bibinfo{pages}{88}.
\newblock


\bibitem[Caldwell(2020)]%
        {n1}
\bibfield{author}{\bibinfo{person}{Craig Caldwell}.} \bibinfo{year}{2020}\natexlab{}.
\newblock \showarticletitle{What we talk about, when we talk about story}.
\newblock In \bibinfo{booktitle}{\emph{ACM SIGGRAPH 2020 Courses}}. \bibinfo{pages}{1--15}.
\newblock


\bibitem[Callahan(2019)]%
        {n6}
\bibfield{author}{\bibinfo{person}{Monica Callahan}.} \bibinfo{year}{2019}\natexlab{}.
\newblock \bibinfo{title}{Presenting foreshadowing}.
\newblock
\newblock


\bibitem[Catmull(2008)]%
        {n14}
\bibfield{author}{\bibinfo{person}{Edwin Catmull}.} \bibinfo{year}{2008}\natexlab{}.
\newblock \bibinfo{booktitle}{\emph{How Pixar fosters collective creativity}}.
\newblock \bibinfo{publisher}{Harvard Business School Publishing Boston, MA}.
\newblock


\bibitem[Chan et~al\mbox{.}({[n.\,d.]})]%
        {social8}
\bibfield{author}{\bibinfo{person}{Chi-Min Chan}, \bibinfo{person}{Weize Chen}, \bibinfo{person}{Yusheng Su}, \bibinfo{person}{Jianxuan Yu}, \bibinfo{person}{Wei Xue}, \bibinfo{person}{Shanghang Zhang}, \bibinfo{person}{Jie Fu}, {and} \bibinfo{person}{Zhiyuan Liu}.} \bibinfo{year}{[n.\,d.]}\natexlab{}.
\newblock \showarticletitle{ChatEval: Towards Better LLM-based Evaluators through Multi-Agent Debate}. In \bibinfo{booktitle}{\emph{The Twelfth International Conference on Learning Representations}}.
\newblock


\bibitem[Chen et~al\mbox{.}(2023b)]%
        {m10}
\bibfield{author}{\bibinfo{person}{Dake Chen}, \bibinfo{person}{Hanbin Wang}, \bibinfo{person}{Yunhao Huo}, \bibinfo{person}{Yuzhao Li}, {and} \bibinfo{person}{Haoyang Zhang}.} \bibinfo{year}{2023}\natexlab{b}.
\newblock \showarticletitle{Gamegpt: Multi-agent collaborative framework for game development}.
\newblock \bibinfo{journal}{\emph{arXiv preprint arXiv:2310.08067}} (\bibinfo{year}{2023}).
\newblock


\bibitem[Chen(2007)]%
        {social10}
\bibfield{author}{\bibinfo{person}{Jenova Chen}.} \bibinfo{year}{2007}\natexlab{}.
\newblock \showarticletitle{Flow in games (and everything else)}.
\newblock \bibinfo{journal}{\emph{Commun. ACM}} \bibinfo{volume}{50}, \bibinfo{number}{4} (\bibinfo{year}{2007}), \bibinfo{pages}{31--34}.
\newblock


\bibitem[Chen et~al\mbox{.}(2022)]%
        {v12}
\bibfield{author}{\bibinfo{person}{Nuo Chen}, \bibinfo{person}{Yan Wang}, \bibinfo{person}{Haiyun Jiang}, \bibinfo{person}{Deng Cai}, \bibinfo{person}{Yuhan Li}, \bibinfo{person}{Ziyang Chen}, \bibinfo{person}{Longyue Wang}, {and} \bibinfo{person}{Jia Li}.} \bibinfo{year}{2022}\natexlab{}.
\newblock \showarticletitle{Large Language Models Meet Harry Potter: A Bilingual Dataset for Aligning Dialogue Agents with Characters}.
\newblock \bibinfo{journal}{\emph{arXiv preprint arXiv:2211.06869}} (\bibinfo{year}{2022}).
\newblock


\bibitem[Chen et~al\mbox{.}(2023a)]%
        {m23}
\bibfield{author}{\bibinfo{person}{Weize Chen}, \bibinfo{person}{Yusheng Su}, \bibinfo{person}{Jingwei Zuo}, \bibinfo{person}{Cheng Yang}, \bibinfo{person}{Chenfei Yuan}, \bibinfo{person}{Chen Qian}, \bibinfo{person}{Chi-Min Chan}, \bibinfo{person}{Yujia Qin}, \bibinfo{person}{Yaxi Lu}, \bibinfo{person}{Ruobing Xie}, {et~al\mbox{.}}} \bibinfo{year}{2023}\natexlab{a}.
\newblock \showarticletitle{Agentverse: Facilitating multi-agent collaboration and exploring emergent behaviors in agents}.
\newblock \bibinfo{journal}{\emph{arXiv preprint arXiv:2308.10848}} (\bibinfo{year}{2023}).
\newblock


\bibitem[Cheng et~al\mbox{.}(2024)]%
        {m33}
\bibfield{author}{\bibinfo{person}{Yuheng Cheng}, \bibinfo{person}{Ceyao Zhang}, \bibinfo{person}{Zhengwen Zhang}, \bibinfo{person}{Xiangrui Meng}, \bibinfo{person}{Sirui Hong}, \bibinfo{person}{Wenhao Li}, \bibinfo{person}{Zihao Wang}, \bibinfo{person}{Zekai Wang}, \bibinfo{person}{Feng Yin}, \bibinfo{person}{Junhua Zhao}, {et~al\mbox{.}}} \bibinfo{year}{2024}\natexlab{}.
\newblock \showarticletitle{Exploring Large Language Model based Intelligent Agents: Definitions, Methods, and Prospects}.
\newblock \bibinfo{journal}{\emph{arXiv preprint arXiv:2401.03428}} (\bibinfo{year}{2024}).
\newblock


\bibitem[Choi et~al\mbox{.}(2021)]%
        {v31}
\bibfield{author}{\bibinfo{person}{Seongho Choi}, \bibinfo{person}{Kyoung-Woon On}, \bibinfo{person}{Yu-Jung Heo}, \bibinfo{person}{Ahjeong Seo}, \bibinfo{person}{Youwon Jang}, \bibinfo{person}{Minsu Lee}, {and} \bibinfo{person}{Byoung-Tak Zhang}.} \bibinfo{year}{2021}\natexlab{}.
\newblock \showarticletitle{Dramaqa: Character-centered video story understanding with hierarchical qa}. In \bibinfo{booktitle}{\emph{Proceedings of the AAAI Conference on Artificial Intelligence}}, Vol.~\bibinfo{volume}{35}. \bibinfo{pages}{1166--1174}.
\newblock


\bibitem[Chopra et~al\mbox{.}(2021)]%
        {n16}
\bibfield{author}{\bibinfo{person}{Bhavya Chopra}, \bibinfo{person}{Khushali Verma}, \bibinfo{person}{Sonali Singhal}, {and} \bibinfo{person}{Utsav Singla}.} \bibinfo{year}{2021}\natexlab{}.
\newblock \showarticletitle{Reality Tales: Facilitating User-Character Interaction with Immersive Storytelling}. In \bibinfo{booktitle}{\emph{Extended Abstracts of the 2021 CHI Conference on Human Factors in Computing Systems}}. \bibinfo{pages}{1--7}.
\newblock


\bibitem[Chuang et~al\mbox{.}(2024)]%
        {social7}
\bibfield{author}{\bibinfo{person}{Yun-Shiuan Chuang}, \bibinfo{person}{Nikunj Harlalka}, \bibinfo{person}{Siddharth Suresh}, \bibinfo{person}{Agam Goyal}, \bibinfo{person}{Robert Hawkins}, \bibinfo{person}{Sijia Yang}, \bibinfo{person}{Dhavan Shah}, \bibinfo{person}{Junjie Hu}, {and} \bibinfo{person}{Timothy~T Rogers}.} \bibinfo{year}{2024}\natexlab{}.
\newblock \showarticletitle{The Wisdom of Partisan Crowds: Comparing Collective Intelligence in Humans and LLM-based Agents}. In \bibinfo{booktitle}{\emph{Proceedings of the Annual Meeting of the Cognitive Science Society}}, Vol.~\bibinfo{volume}{46}.
\newblock


\bibitem[Czikszentmihalyi(1990)]%
        {social11}
\bibfield{author}{\bibinfo{person}{Mihaly Czikszentmihalyi}.} \bibinfo{year}{1990}\natexlab{}.
\newblock \bibinfo{booktitle}{\emph{Flow: The psychology of optimal experience}}.
\newblock \bibinfo{publisher}{New York: Harper \& Row}.
\newblock


\bibitem[Dahdal(2020)]%
        {n18}
\bibfield{author}{\bibinfo{person}{Sohail Dahdal}.} \bibinfo{year}{2020}\natexlab{}.
\newblock \showarticletitle{The illusive ludonarrativity and the problem with emergent interactive storytelling models in interactive movies}.
\newblock \bibinfo{journal}{\emph{Journal of Digital Media \& Interaction}} \bibinfo{volume}{3}, \bibinfo{number}{6} (\bibinfo{year}{2020}), \bibinfo{pages}{17--33}.
\newblock


\bibitem[Dong et~al\mbox{.}(2022)]%
        {m29}
\bibfield{author}{\bibinfo{person}{Qingxiu Dong}, \bibinfo{person}{Lei Li}, \bibinfo{person}{Damai Dai}, \bibinfo{person}{Ce Zheng}, \bibinfo{person}{Zhiyong Wu}, \bibinfo{person}{Baobao Chang}, \bibinfo{person}{Xu Sun}, \bibinfo{person}{Jingjing Xu}, {and} \bibinfo{person}{Zhifang Sui}.} \bibinfo{year}{2022}\natexlab{}.
\newblock \showarticletitle{A survey on in-context learning}.
\newblock \bibinfo{journal}{\emph{arXiv preprint arXiv:2301.00234}} (\bibinfo{year}{2022}).
\newblock


\bibitem[Dong et~al\mbox{.}(2023)]%
        {m1}
\bibfield{author}{\bibinfo{person}{Yihong Dong}, \bibinfo{person}{Xue Jiang}, \bibinfo{person}{Zhi Jin}, {and} \bibinfo{person}{Ge Li}.} \bibinfo{year}{2023}\natexlab{}.
\newblock \showarticletitle{Self-collaboration code generation via chatgpt}.
\newblock \bibinfo{journal}{\emph{arXiv preprint arXiv:2304.07590}} (\bibinfo{year}{2023}).
\newblock


\bibitem[Durante et~al\mbox{.}(2024)]%
        {m16}
\bibfield{author}{\bibinfo{person}{Zane Durante}, \bibinfo{person}{Qiuyuan Huang}, \bibinfo{person}{Naoki Wake}, \bibinfo{person}{Ran Gong}, \bibinfo{person}{Jae~Sung Park}, \bibinfo{person}{Bidipta Sarkar}, \bibinfo{person}{Rohan Taori}, \bibinfo{person}{Yusuke Noda}, \bibinfo{person}{Demetri Terzopoulos}, \bibinfo{person}{Yejin Choi}, {et~al\mbox{.}}} \bibinfo{year}{2024}\natexlab{}.
\newblock \showarticletitle{Agent ai: Surveying the horizons of multimodal interaction}.
\newblock \bibinfo{journal}{\emph{arXiv preprint arXiv:2401.03568}} (\bibinfo{year}{2024}).
\newblock


\bibitem[Gao et~al\mbox{.}(2023)]%
        {t3}
\bibfield{author}{\bibinfo{person}{Yunfan Gao}, \bibinfo{person}{Yun Xiong}, \bibinfo{person}{Xinyu Gao}, \bibinfo{person}{Kangxiang Jia}, \bibinfo{person}{Jinliu Pan}, \bibinfo{person}{Yuxi Bi}, \bibinfo{person}{Yi Dai}, \bibinfo{person}{Jiawei Sun}, {and} \bibinfo{person}{Haofen Wang}.} \bibinfo{year}{2023}\natexlab{}.
\newblock \showarticletitle{Retrieval-augmented generation for large language models: A survey}.
\newblock \bibinfo{journal}{\emph{arXiv preprint arXiv:2312.10997}} (\bibinfo{year}{2023}).
\newblock


\bibitem[Gerv{\'a}s(2009)]%
        {n9}
\bibfield{author}{\bibinfo{person}{Pablo Gerv{\'a}s}.} \bibinfo{year}{2009}\natexlab{}.
\newblock \showarticletitle{Computational approaches to storytelling and creativity}.
\newblock \bibinfo{journal}{\emph{AI Magazine}} \bibinfo{volume}{30}, \bibinfo{number}{3} (\bibinfo{year}{2009}), \bibinfo{pages}{49--49}.
\newblock


\bibitem[Guo et~al\mbox{.}(2024)]%
        {t4}
\bibfield{author}{\bibinfo{person}{Taicheng Guo}, \bibinfo{person}{Xiuying Chen}, \bibinfo{person}{Yaqi Wang}, \bibinfo{person}{Ruidi Chang}, \bibinfo{person}{Shichao Pei}, \bibinfo{person}{Nitesh~V Chawla}, \bibinfo{person}{Olaf Wiest}, {and} \bibinfo{person}{Xiangliang Zhang}.} \bibinfo{year}{2024}\natexlab{}.
\newblock \showarticletitle{Large language model based multi-agents: A survey of progress and challenges}.
\newblock \bibinfo{journal}{\emph{arXiv preprint arXiv:2402.01680}} (\bibinfo{year}{2024}).
\newblock


\bibitem[Hang and Azahari(2023)]%
        {n19}
\bibfield{author}{\bibinfo{person}{Sun Hang} {and} \bibinfo{person}{Mustaffa~Halabi Azahari}.} \bibinfo{year}{2023}\natexlab{}.
\newblock \showarticletitle{Interactive movie in digital era: A systematic literature review}.
\newblock  (\bibinfo{year}{2023}).
\newblock


\bibitem[Hong et~al\mbox{.}(2023)]%
        {m5}
\bibfield{author}{\bibinfo{person}{Sirui Hong}, \bibinfo{person}{Xiawu Zheng}, \bibinfo{person}{Jonathan Chen}, \bibinfo{person}{Yuheng Cheng}, \bibinfo{person}{Jinlin Wang}, \bibinfo{person}{Ceyao Zhang}, \bibinfo{person}{Zili Wang}, \bibinfo{person}{Steven Ka~Shing Yau}, \bibinfo{person}{Zijuan Lin}, \bibinfo{person}{Liyang Zhou}, {et~al\mbox{.}}} \bibinfo{year}{2023}\natexlab{}.
\newblock \showarticletitle{Metagpt: Meta programming for multi-agent collaborative framework}.
\newblock \bibinfo{journal}{\emph{arXiv preprint arXiv:2308.00352}} (\bibinfo{year}{2023}).
\newblock


\bibitem[Hua et~al\mbox{.}(2023)]%
        {m19}
\bibfield{author}{\bibinfo{person}{Wenyue Hua}, \bibinfo{person}{Lizhou Fan}, \bibinfo{person}{Lingyao Li}, \bibinfo{person}{Kai Mei}, \bibinfo{person}{Jianchao Ji}, \bibinfo{person}{Yingqiang Ge}, \bibinfo{person}{Libby Hemphill}, {and} \bibinfo{person}{Yongfeng Zhang}.} \bibinfo{year}{2023}\natexlab{}.
\newblock \showarticletitle{War and peace (waragent): Large language model-based multi-agent simulation of world wars}.
\newblock \bibinfo{journal}{\emph{arXiv preprint arXiv:2311.17227}} (\bibinfo{year}{2023}).
\newblock


\bibitem[Huang et~al\mbox{.}(2024)]%
        {m24}
\bibfield{author}{\bibinfo{person}{Qiuyuan Huang}, \bibinfo{person}{Naoki Wake}, \bibinfo{person}{Bidipta Sarkar}, \bibinfo{person}{Zane Durante}, \bibinfo{person}{Ran Gong}, \bibinfo{person}{Rohan Taori}, \bibinfo{person}{Yusuke Noda}, \bibinfo{person}{Demetri Terzopoulos}, \bibinfo{person}{Noboru Kuno}, \bibinfo{person}{Ade Famoti}, {et~al\mbox{.}}} \bibinfo{year}{2024}\natexlab{}.
\newblock \showarticletitle{Position Paper: Agent AI Towards a Holistic Intelligence}.
\newblock \bibinfo{journal}{\emph{arXiv preprint arXiv:2403.00833}} (\bibinfo{year}{2024}).
\newblock


\bibitem[Huang et~al\mbox{.}(2020)]%
        {v8}
\bibfield{author}{\bibinfo{person}{Qingqiu Huang}, \bibinfo{person}{Yu Xiong}, \bibinfo{person}{Anyi Rao}, \bibinfo{person}{Jiaze Wang}, {and} \bibinfo{person}{Dahua Lin}.} \bibinfo{year}{2020}\natexlab{}.
\newblock \showarticletitle{Movienet: A holistic dataset for movie understanding}. In \bibinfo{booktitle}{\emph{Computer Vision--ECCV 2020: 16th European Conference, Glasgow, UK, August 23--28, 2020, Proceedings, Part IV 16}}. Springer, \bibinfo{pages}{709--727}.
\newblock


\bibitem[Huang(2024)]%
        {m22}
\bibfield{author}{\bibinfo{person}{Yu Huang}.} \bibinfo{year}{2024}\natexlab{}.
\newblock \showarticletitle{Levels of AI Agents: from Rules to Large Language Models}.
\newblock \bibinfo{journal}{\emph{arXiv preprint arXiv:2405.06643}} (\bibinfo{year}{2024}).
\newblock


\bibitem[Hussain et~al\mbox{.}(2020)]%
        {n2}
\bibfield{author}{\bibinfo{person}{Afzal Hussain}, \bibinfo{person}{Haad Shakeel}, \bibinfo{person}{Faizan Hussain}, \bibinfo{person}{Nasir Uddin}, {and} \bibinfo{person}{Turab~Latif Ghouri}.} \bibinfo{year}{2020}\natexlab{}.
\newblock \showarticletitle{Unity game development engine: A technical survey}.
\newblock \bibinfo{journal}{\emph{Univ. Sindh J. Inf. Commun. Technol}} \bibinfo{volume}{4}, \bibinfo{number}{2} (\bibinfo{year}{2020}), \bibinfo{pages}{73--81}.
\newblock


\bibitem[Liang et~al\mbox{.}(2023)]%
        {m20}
\bibfield{author}{\bibinfo{person}{Tian Liang}, \bibinfo{person}{Zhiwei He}, \bibinfo{person}{Jen-tes Huang}, \bibinfo{person}{Wenxuan Wang}, \bibinfo{person}{Wenxiang Jiao}, \bibinfo{person}{Rui Wang}, \bibinfo{person}{Yujiu Yang}, \bibinfo{person}{Zhaopeng Tu}, \bibinfo{person}{Shuming Shi}, {and} \bibinfo{person}{Xing Wang}.} \bibinfo{year}{2023}\natexlab{}.
\newblock \showarticletitle{Leveraging Word Guessing Games to Assess the Intelligence of Large Language Models}.
\newblock \bibinfo{journal}{\emph{arXiv preprint arXiv:2310.20499}} (\bibinfo{year}{2023}).
\newblock


\bibitem[Likert(1932)]%
        {social6}
\bibfield{author}{\bibinfo{person}{Rensis Likert}.} \bibinfo{year}{1932}\natexlab{}.
\newblock \showarticletitle{A technique for the measurement of attitudes.}
\newblock \bibinfo{journal}{\emph{Archives of psychology}} (\bibinfo{year}{1932}).
\newblock


\bibitem[Lin et~al\mbox{.}(2024)]%
        {m36}
\bibfield{author}{\bibinfo{person}{Shuhang Lin}, \bibinfo{person}{Wenyue Hua}, \bibinfo{person}{Lingyao Li}, \bibinfo{person}{Che-Jui Chang}, \bibinfo{person}{Lizhou Fan}, \bibinfo{person}{Jianchao Ji}, \bibinfo{person}{Hang Hua}, \bibinfo{person}{Mingyu Jin}, \bibinfo{person}{Jiebo Luo}, {and} \bibinfo{person}{Yongfeng Zhang}.} \bibinfo{year}{2024}\natexlab{}.
\newblock \showarticletitle{BattleAgent: Multi-modal Dynamic Emulation on Historical Battles to Complement Historical Analysis}.
\newblock \bibinfo{journal}{\emph{arXiv preprint arXiv:2404.15532}} (\bibinfo{year}{2024}).
\newblock


\bibitem[Liu et~al\mbox{.}(2020)]%
        {v35}
\bibfield{author}{\bibinfo{person}{Chang Liu}, \bibinfo{person}{Armin Shmilovici}, {and} \bibinfo{person}{Mark Last}.} \bibinfo{year}{2020}\natexlab{}.
\newblock \showarticletitle{Towards story-based classification of movie scenes}.
\newblock \bibinfo{journal}{\emph{PloS one}} \bibinfo{volume}{15}, \bibinfo{number}{2} (\bibinfo{year}{2020}), \bibinfo{pages}{e0228579}.
\newblock


\bibitem[London et~al\mbox{.}(2023)]%
        {social3}
\bibfield{author}{\bibinfo{person}{Manuel London}, \bibinfo{person}{Valerie~I Sessa}, {and} \bibinfo{person}{Loren~A Shelley}.} \bibinfo{year}{2023}\natexlab{}.
\newblock \showarticletitle{Developing self-awareness: Learning processes for self-and interpersonal growth}.
\newblock \bibinfo{journal}{\emph{Annual Review of Organizational Psychology and Organizational Behavior}}  \bibinfo{volume}{10} (\bibinfo{year}{2023}), \bibinfo{pages}{261--288}.
\newblock


\bibitem[Lu et~al\mbox{.}(2024)]%
        {v40}
\bibfield{author}{\bibinfo{person}{Jiaying Lu}, \bibinfo{person}{Bo Pan}, \bibinfo{person}{Jieyi Chen}, \bibinfo{person}{Yingchaojie Feng}, \bibinfo{person}{Jingyuan Hu}, \bibinfo{person}{Yuchen Peng}, {and} \bibinfo{person}{Wei Chen}.} \bibinfo{year}{2024}\natexlab{}.
\newblock \showarticletitle{AgentLens: Visual Analysis for Agent Behaviors in LLM-based Autonomous Systems}.
\newblock \bibinfo{journal}{\emph{arXiv preprint arXiv:2402.08995}} (\bibinfo{year}{2024}).
\newblock


\bibitem[Ma(2023)]%
        {n17}
\bibfield{author}{\bibinfo{person}{Lynshao~Celina Ma}.} \bibinfo{year}{2023}\natexlab{}.
\newblock \showarticletitle{Enriched Story Experiences with a New Video Interaction Model}.
\newblock  (\bibinfo{year}{2023}).
\newblock


\bibitem[Ma et~al\mbox{.}(2023)]%
        {m6}
\bibfield{author}{\bibinfo{person}{Weiyu Ma}, \bibinfo{person}{Qirui Mi}, \bibinfo{person}{Xue Yan}, \bibinfo{person}{Yuqiao Wu}, \bibinfo{person}{Runji Lin}, \bibinfo{person}{Haifeng Zhang}, {and} \bibinfo{person}{Jun Wang}.} \bibinfo{year}{2023}\natexlab{}.
\newblock \showarticletitle{Large language models play starcraft ii: Benchmarks and a chain of summarization approach}.
\newblock \bibinfo{journal}{\emph{arXiv preprint arXiv:2312.11865}} (\bibinfo{year}{2023}).
\newblock


\bibitem[March et~al\mbox{.}(2023)]%
        {m8}
\bibfield{author}{\bibinfo{person}{David March}, \bibinfo{person}{Julia M{\'u}gica}, \bibinfo{person}{Ezequiel~E Ferrero}, {and} \bibinfo{person}{M~Carmen Miguel}.} \bibinfo{year}{2023}\natexlab{}.
\newblock \showarticletitle{Honeybee-like collective decision making in a kilobot swarm}.
\newblock \bibinfo{journal}{\emph{arXiv preprint arXiv:2310.15592}} (\bibinfo{year}{2023}).
\newblock


\bibitem[Masterman et~al\mbox{.}(2024)]%
        {m18}
\bibfield{author}{\bibinfo{person}{Tula Masterman}, \bibinfo{person}{Sandi Besen}, \bibinfo{person}{Mason Sawtell}, {and} \bibinfo{person}{Alex Chao}.} \bibinfo{year}{2024}\natexlab{}.
\newblock \showarticletitle{The Landscape of Emerging AI Agent Architectures for Reasoning, Planning, and Tool Calling: A Survey}.
\newblock \bibinfo{journal}{\emph{arXiv preprint arXiv:2404.11584}} (\bibinfo{year}{2024}).
\newblock


\bibitem[Maziarczyk(2023)]%
        {n21}
\bibfield{author}{\bibinfo{person}{Grzegorz Maziarczyk}.} \bibinfo{year}{2023}\natexlab{}.
\newblock \showarticletitle{“The road not taken”: an interactive film between narrative and database}.
\newblock \bibinfo{journal}{\emph{New Review of Hypermedia and Multimedia}} \bibinfo{volume}{29}, \bibinfo{number}{1} (\bibinfo{year}{2023}), \bibinfo{pages}{56--71}.
\newblock


\bibitem[McKee(1997)]%
        {n12}
\bibfield{author}{\bibinfo{person}{Robert McKee}.} \bibinfo{year}{1997}\natexlab{}.
\newblock \showarticletitle{Substance, structure, style, and the principles of screenwriting}.
\newblock \bibinfo{journal}{\emph{Alba Editorial}} (\bibinfo{year}{1997}).
\newblock


\bibitem[Milgram(1963)]%
        {social5}
\bibfield{author}{\bibinfo{person}{Stanley Milgram}.} \bibinfo{year}{1963}\natexlab{}.
\newblock \showarticletitle{Behavioral study of obedience.}
\newblock \bibinfo{journal}{\emph{The Journal of abnormal and social psychology}} \bibinfo{volume}{67}, \bibinfo{number}{4} (\bibinfo{year}{1963}), \bibinfo{pages}{371}.
\newblock


\bibitem[Milgram and Gudehus(1974)]%
        {social2}
\bibfield{author}{\bibinfo{person}{Stanley Milgram} {and} \bibinfo{person}{Christian Gudehus}.} \bibinfo{year}{1974}\natexlab{}.
\newblock \bibinfo{title}{Obedience to authority}.
\newblock
\newblock


\bibitem[Minaee et~al\mbox{.}(2024)]%
        {m28}
\bibfield{author}{\bibinfo{person}{Shervin Minaee}, \bibinfo{person}{Tomas Mikolov}, \bibinfo{person}{Narjes Nikzad}, \bibinfo{person}{Meysam Chenaghlu}, \bibinfo{person}{Richard Socher}, \bibinfo{person}{Xavier Amatriain}, {and} \bibinfo{person}{Jianfeng Gao}.} \bibinfo{year}{2024}\natexlab{}.
\newblock \showarticletitle{Large language models: A survey}.
\newblock \bibinfo{journal}{\emph{arXiv preprint arXiv:2402.06196}} (\bibinfo{year}{2024}).
\newblock


\bibitem[Murray(2018)]%
        {n22}
\bibfield{author}{\bibinfo{person}{Janet~H Murray}.} \bibinfo{year}{2018}\natexlab{}.
\newblock \showarticletitle{Research into interactive digital narrative: a kaleidoscopic view}. In \bibinfo{booktitle}{\emph{Interactive Storytelling: 11th International Conference on Interactive Digital Storytelling, ICIDS 2018, Dublin, Ireland, December 5--8, 2018, Proceedings 11}}. Springer, \bibinfo{pages}{3--17}.
\newblock


\bibitem[Nan et~al\mbox{.}(2015)]%
        {v33}
\bibfield{author}{\bibinfo{person}{Chang-Jun Nan}, \bibinfo{person}{Kyung-Min Kim}, {and} \bibinfo{person}{Byoung-Tak Zhang}.} \bibinfo{year}{2015}\natexlab{}.
\newblock \showarticletitle{Social network analysis of TV drama characters via deep concept hierarchies}. In \bibinfo{booktitle}{\emph{Proceedings of the 2015 IEEE/ACM International Conference on Advances in Social Networks Analysis and Mining 2015}}. \bibinfo{pages}{831--836}.
\newblock


\bibitem[Park et~al\mbox{.}(2023)]%
        {m9}
\bibfield{author}{\bibinfo{person}{Joon~Sung Park}, \bibinfo{person}{Joseph O'Brien}, \bibinfo{person}{Carrie~Jun Cai}, \bibinfo{person}{Meredith~Ringel Morris}, \bibinfo{person}{Percy Liang}, {and} \bibinfo{person}{Michael~S Bernstein}.} \bibinfo{year}{2023}\natexlab{}.
\newblock \showarticletitle{Generative agents: Interactive simulacra of human behavior}. In \bibinfo{booktitle}{\emph{Proceedings of the 36th Annual ACM Symposium on User Interface Software and Technology}}. \bibinfo{pages}{1--22}.
\newblock


\bibitem[Pearce(1994)]%
        {n26}
\bibfield{author}{\bibinfo{person}{Celia Pearce}.} \bibinfo{year}{1994}\natexlab{}.
\newblock \showarticletitle{The ins \& outs of non-linear storytelling}.
\newblock \bibinfo{journal}{\emph{ACM SIGGRAPH Computer Graphics}} \bibinfo{volume}{28}, \bibinfo{number}{2} (\bibinfo{year}{1994}), \bibinfo{pages}{100--101}.
\newblock


\bibitem[Qian et~al\mbox{.}(2024)]%
        {m12}
\bibfield{author}{\bibinfo{person}{Chen Qian}, \bibinfo{person}{Wei Liu}, \bibinfo{person}{Hongzhang Liu}, \bibinfo{person}{Nuo Chen}, \bibinfo{person}{Yufan Dang}, \bibinfo{person}{Jiahao Li}, \bibinfo{person}{Cheng Yang}, \bibinfo{person}{Weize Chen}, \bibinfo{person}{Yusheng Su}, \bibinfo{person}{Xin Cong}, {et~al\mbox{.}}} \bibinfo{year}{2024}\natexlab{}.
\newblock \showarticletitle{Chatdev: Communicative agents for software development}. In \bibinfo{booktitle}{\emph{Proceedings of the 62nd Annual Meeting of the Association for Computational Linguistics (Volume 1: Long Papers)}}. \bibinfo{pages}{15174--15186}.
\newblock


\bibitem[Ranade and Joshi(2023)]%
        {v15}
\bibfield{author}{\bibinfo{person}{Priyanka Ranade} {and} \bibinfo{person}{Anupam Joshi}.} \bibinfo{year}{2023}\natexlab{}.
\newblock \showarticletitle{FABULA: Intelligence Report Generation Using Retrieval-Augmented Narrative Construction}.
\newblock \bibinfo{journal}{\emph{arXiv preprint arXiv:2310.13848}} (\bibinfo{year}{2023}).
\newblock


\bibitem[Riedl and Young(2010)]%
        {n7}
\bibfield{author}{\bibinfo{person}{Mark~O Riedl} {and} \bibinfo{person}{Robert~Michael Young}.} \bibinfo{year}{2010}\natexlab{}.
\newblock \showarticletitle{Narrative planning: Balancing plot and character}.
\newblock \bibinfo{journal}{\emph{Journal of Artificial Intelligence Research}}  \bibinfo{volume}{39} (\bibinfo{year}{2010}), \bibinfo{pages}{217--268}.
\newblock


\bibitem[Roth and Koenitz(2019)]%
        {n20}
\bibfield{author}{\bibinfo{person}{Christian Roth} {and} \bibinfo{person}{Hartmut Koenitz}.} \bibinfo{year}{2019}\natexlab{}.
\newblock \showarticletitle{Bandersnatch, yea or nay? Reception and user experience of an interactive digital narrative video}. In \bibinfo{booktitle}{\emph{Proceedings of the 2019 ACM International Conference on Interactive Experiences for TV and Online Video}}. \bibinfo{pages}{247--254}.
\newblock


\bibitem[Ryan(2015)]%
        {n15}
\bibfield{author}{\bibinfo{person}{Marie-Laure Ryan}.} \bibinfo{year}{2015}\natexlab{}.
\newblock \bibinfo{booktitle}{\emph{Narrative as virtual reality 2: Revisiting immersion and interactivity in literature and electronic media}}.
\newblock \bibinfo{publisher}{JHU press}.
\newblock


\bibitem[Shannon(1948)]%
        {m32}
\bibfield{author}{\bibinfo{person}{Claude~Elwood Shannon}.} \bibinfo{year}{1948}\natexlab{}.
\newblock \showarticletitle{A mathematical theory of communication}.
\newblock \bibinfo{journal}{\emph{The Bell system technical journal}} \bibinfo{volume}{27}, \bibinfo{number}{3} (\bibinfo{year}{1948}), \bibinfo{pages}{379--423}.
\newblock


\bibitem[Shao et~al\mbox{.}(2023)]%
        {v22}
\bibfield{author}{\bibinfo{person}{Yunfan Shao}, \bibinfo{person}{Linyang Li}, \bibinfo{person}{Junqi Dai}, {and} \bibinfo{person}{Xipeng Qiu}.} \bibinfo{year}{2023}\natexlab{}.
\newblock \bibinfo{title}{Character-LLM: A Trainable Agent for Role-Playing}.
\newblock
\newblock
\showeprint[arxiv]{2310.10158}~[cs.CL]


\bibitem[Silvia and Duval(2001)]%
        {social1}
\bibfield{author}{\bibinfo{person}{Paul~J Silvia} {and} \bibinfo{person}{T~Shelley Duval}.} \bibinfo{year}{2001}\natexlab{}.
\newblock \showarticletitle{Objective self-awareness theory: Recent progress and enduring problems}.
\newblock \bibinfo{journal}{\emph{Personality and social psychology review}} \bibinfo{volume}{5}, \bibinfo{number}{3} (\bibinfo{year}{2001}), \bibinfo{pages}{230--241}.
\newblock


\bibitem[Somvanshi(2023)]%
        {v37}
\bibfield{author}{\bibinfo{person}{Dushyant Somvanshi}.} \bibinfo{year}{2023}\natexlab{}.
\newblock \showarticletitle{Audio-Visual Storytelling Through Immersive Media and its Impact: A Critical Literature Review}.
\newblock \bibinfo{journal}{\emph{European Chemical Bulletin}} (\bibinfo{year}{2023}).
\newblock


\bibitem[Sun et~al\mbox{.}(2022)]%
        {v39}
\bibfield{author}{\bibinfo{person}{Mengdi Sun}, \bibinfo{person}{Ligan Cai}, \bibinfo{person}{Weiwei Cui}, \bibinfo{person}{Yanqiu Wu}, \bibinfo{person}{Yang Shi}, {and} \bibinfo{person}{Nan Cao}.} \bibinfo{year}{2022}\natexlab{}.
\newblock \showarticletitle{Erato: Cooperative data story editing via fact interpolation}.
\newblock \bibinfo{journal}{\emph{IEEE Transactions on Visualization and Computer Graphics}} \bibinfo{volume}{29}, \bibinfo{number}{1} (\bibinfo{year}{2022}), \bibinfo{pages}{983--993}.
\newblock


\bibitem[Tanahashi and Ma(2012)]%
        {v36}
\bibfield{author}{\bibinfo{person}{Yuzuru Tanahashi} {and} \bibinfo{person}{Kwan-Liu Ma}.} \bibinfo{year}{2012}\natexlab{}.
\newblock \showarticletitle{Design considerations for optimizing storyline visualizations}.
\newblock \bibinfo{journal}{\emph{IEEE Transactions on Visualization and Computer Graphics}} \bibinfo{volume}{18}, \bibinfo{number}{12} (\bibinfo{year}{2012}), \bibinfo{pages}{2679--2688}.
\newblock


\bibitem[Ursu et~al\mbox{.}(2008)]%
        {n25}
\bibfield{author}{\bibinfo{person}{Marian~F Ursu}, \bibinfo{person}{Maureen Thomas}, \bibinfo{person}{Ian Kegel}, \bibinfo{person}{Doug Williams}, \bibinfo{person}{Mika Tuomola}, \bibinfo{person}{Inger Lindstedt}, \bibinfo{person}{Terence Wright}, \bibinfo{person}{Andra Leurdijk}, \bibinfo{person}{Vilmos Zsombori}, \bibinfo{person}{Julia Sussner}, {et~al\mbox{.}}} \bibinfo{year}{2008}\natexlab{}.
\newblock \showarticletitle{Interactive TV narratives: Opportunities, progress, and challenges}.
\newblock \bibinfo{journal}{\emph{ACM Transactions on Multimedia Computing, Communications, and Applications (TOMM)}} \bibinfo{volume}{4}, \bibinfo{number}{4} (\bibinfo{year}{2008}), \bibinfo{pages}{1--39}.
\newblock


\bibitem[Verma et~al\mbox{.}(2023)]%
        {m30}
\bibfield{author}{\bibinfo{person}{Kirti Verma}, \bibinfo{person}{Sateesh Kourav}, \bibinfo{person}{Mukul Jangid}, \bibinfo{person}{Uma Sahu}, {and} \bibinfo{person}{Neeraj Shivhare}.} \bibinfo{year}{2023}\natexlab{}.
\newblock \showarticletitle{Research on Finite State Machine and Its Real Life Time Applications}.
\newblock  (\bibinfo{date}{12} \bibinfo{year}{2023}).
\newblock


\bibitem[Vicol et~al\mbox{.}(2018)]%
        {v34}
\bibfield{author}{\bibinfo{person}{Paul Vicol}, \bibinfo{person}{Makarand Tapaswi}, \bibinfo{person}{Lluis Castrejon}, {and} \bibinfo{person}{Sanja Fidler}.} \bibinfo{year}{2018}\natexlab{}.
\newblock \showarticletitle{Moviegraphs: Towards understanding human-centric situations from videos}. In \bibinfo{booktitle}{\emph{Proceedings of the IEEE conference on computer vision and pattern recognition}}. \bibinfo{pages}{8581--8590}.
\newblock


\bibitem[Wang et~al\mbox{.}({[n.\,d.]})]%
        {m17}
\bibfield{author}{\bibinfo{person}{Guanzhi Wang}, \bibinfo{person}{Yuqi Xie}, \bibinfo{person}{Yunfan Jiang}, \bibinfo{person}{Ajay Mandlekar}, \bibinfo{person}{Chaowei Xiao}, \bibinfo{person}{Yuke Zhu}, \bibinfo{person}{Linxi Fan}, {and} \bibinfo{person}{Anima Anandkumar}.} \bibinfo{year}{[n.\,d.]}\natexlab{}.
\newblock \showarticletitle{Voyager: An Open-Ended Embodied Agent with Large Language Models}.
\newblock \bibinfo{journal}{\emph{Transactions on Machine Learning Research}} (\bibinfo{year}{[n.\,d.]}).
\newblock


\bibitem[Wang et~al\mbox{.}(2023)]%
        {m4}
\bibfield{author}{\bibinfo{person}{Zhenhailong Wang}, \bibinfo{person}{Shaoguang Mao}, \bibinfo{person}{Wenshan Wu}, \bibinfo{person}{Tao Ge}, \bibinfo{person}{Furu Wei}, {and} \bibinfo{person}{Heng Ji}.} \bibinfo{year}{2023}\natexlab{}.
\newblock \showarticletitle{Unleashing the emergent cognitive synergy in large language models: A task-solving agent through multi-persona self-collaboration}.
\newblock \bibinfo{journal}{\emph{arXiv preprint arXiv:2307.05300}} (\bibinfo{year}{2023}).
\newblock


\bibitem[Wang et~al\mbox{.}(2024)]%
        {m37}
\bibfield{author}{\bibinfo{person}{Zhan Wang}, \bibinfo{person}{Lin-Ping Yuan}, \bibinfo{person}{Liangwei Wang}, \bibinfo{person}{Bingchuan Jiang}, {and} \bibinfo{person}{Wei Zeng}.} \bibinfo{year}{2024}\natexlab{}.
\newblock \showarticletitle{Virtuwander: Enhancing multi-modal interaction for virtual tour guidance through large language models}. In \bibinfo{booktitle}{\emph{Proceedings of the CHI conference on human factors in computing systems}}. \bibinfo{pages}{1--20}.
\newblock


\bibitem[Weng et~al\mbox{.}(2009)]%
        {v32}
\bibfield{author}{\bibinfo{person}{Chung-Yi Weng}, \bibinfo{person}{Wei-Ta Chu}, {and} \bibinfo{person}{Ja-Ling Wu}.} \bibinfo{year}{2009}\natexlab{}.
\newblock \showarticletitle{Rolenet: Movie analysis from the perspective of social networks}.
\newblock \bibinfo{journal}{\emph{IEEE Transactions on Multimedia}} \bibinfo{volume}{11}, \bibinfo{number}{2} (\bibinfo{year}{2009}), \bibinfo{pages}{256--271}.
\newblock


\bibitem[Werhane(2019)]%
        {n3}
\bibfield{author}{\bibinfo{person}{Patricia~H Werhane}.} \bibinfo{year}{2019}\natexlab{}.
\newblock \showarticletitle{The rashomon effect}.
\newblock \bibinfo{journal}{\emph{Systems thinking and moral imagination: Rethinking Business Ethics with Patricia Werhane}} (\bibinfo{year}{2019}), \bibinfo{pages}{335--343}.
\newblock


\bibitem[Wu et~al\mbox{.}(2023b)]%
        {v38}
\bibfield{author}{\bibinfo{person}{Guande Wu}, \bibinfo{person}{Shunan Guo}, \bibinfo{person}{Jane Hoffswell}, \bibinfo{person}{Gromit Yeuk-Yin Chan}, \bibinfo{person}{Ryan~A Rossi}, {and} \bibinfo{person}{Eunyee Koh}.} \bibinfo{year}{2023}\natexlab{b}.
\newblock \showarticletitle{Socrates: Data Story Generation via Adaptive Machine-Guided Elicitation of User Feedback}.
\newblock \bibinfo{journal}{\emph{IEEE Transactions on Visualization and Computer Graphics}} (\bibinfo{year}{2023}).
\newblock


\bibitem[Wu et~al\mbox{.}(2024)]%
        {m34}
\bibfield{author}{\bibinfo{person}{Minghao Wu}, \bibinfo{person}{Yulin Yuan}, \bibinfo{person}{Gholamreza Haffari}, {and} \bibinfo{person}{Longyue Wang}.} \bibinfo{year}{2024}\natexlab{}.
\newblock \showarticletitle{(Perhaps) Beyond Human Translation: Harnessing Multi-Agent Collaboration for Translating Ultra-Long Literary Texts}.
\newblock \bibinfo{journal}{\emph{arXiv preprint arXiv:2405.11804}} (\bibinfo{year}{2024}).
\newblock


\bibitem[Wu et~al\mbox{.}(2023a)]%
        {m14}
\bibfield{author}{\bibinfo{person}{Qingyun Wu}, \bibinfo{person}{Gagan Bansal}, \bibinfo{person}{Jieyu Zhang}, \bibinfo{person}{Yiran Wu}, \bibinfo{person}{Shaokun Zhang}, \bibinfo{person}{Erkang Zhu}, \bibinfo{person}{Beibin Li}, \bibinfo{person}{Li Jiang}, \bibinfo{person}{Xiaoyun Zhang}, {and} \bibinfo{person}{Chi Wang}.} \bibinfo{year}{2023}\natexlab{a}.
\newblock \showarticletitle{Autogen: Enabling next-gen llm applications via multi-agent conversation framework}.
\newblock \bibinfo{journal}{\emph{arXiv preprint arXiv:2308.08155}} (\bibinfo{year}{2023}).
\newblock


\bibitem[Xi et~al\mbox{.}(2024)]%
        {m25}
\bibfield{author}{\bibinfo{person}{Zhiheng Xi}, \bibinfo{person}{Yiwen Ding}, \bibinfo{person}{Wenxiang Chen}, \bibinfo{person}{Boyang Hong}, \bibinfo{person}{Honglin Guo}, \bibinfo{person}{Junzhe Wang}, \bibinfo{person}{Dingwen Yang}, \bibinfo{person}{Chenyang Liao}, \bibinfo{person}{Xin Guo}, \bibinfo{person}{Wei He}, \bibinfo{person}{Songyang Gao}, \bibinfo{person}{Lu Chen}, \bibinfo{person}{Rui Zheng}, \bibinfo{person}{Yicheng Zou}, \bibinfo{person}{Tao Gui}, \bibinfo{person}{Qi Zhang}, \bibinfo{person}{Xipeng Qiu}, \bibinfo{person}{Xuanjing Huang}, \bibinfo{person}{Zuxuan Wu}, {and} \bibinfo{person}{Yu-Gang Jiang}.} \bibinfo{year}{2024}\natexlab{}.
\newblock \bibinfo{title}{AgentGym: Evolving Large Language Model-based Agents across Diverse Environments}.
\newblock
\newblock
\showeprint[arxiv]{2406.04151}~[cs.AI]


\bibitem[Xiaoliang(2024)]%
        {m31}
\bibfield{author}{\bibinfo{person}{Qi Xiaoliang}.} \bibinfo{year}{2024}\natexlab{}.
\newblock \showarticletitle{Time, Information and Artificial Intelligence}.
\newblock \bibinfo{journal}{\emph{Physics}} \bibinfo{volume}{53}, \bibinfo{number}{6} (\bibinfo{year}{2024}), \bibinfo{pages}{357--367}.
\newblock


\bibitem[Zeng et~al\mbox{.}(2024)]%
        {social9}
\bibfield{author}{\bibinfo{person}{Xingchen Zeng}, \bibinfo{person}{Ziyao Gao}, \bibinfo{person}{Yilin Ye}, {and} \bibinfo{person}{Wei Zeng}.} \bibinfo{year}{2024}\natexlab{}.
\newblock \showarticletitle{IntentTuner: An Interactive Framework for Integrating Human Intentions in Fine-tuning Text-to-Image Generative Models}. In \bibinfo{booktitle}{\emph{Proceedings of the CHI Conference on Human Factors in Computing Systems}}. \bibinfo{pages}{1--18}.
\newblock


\bibitem[Zhao et~al\mbox{.}(2023)]%
        {m35}
\bibfield{author}{\bibinfo{person}{Qinlin Zhao}, \bibinfo{person}{Jindong Wang}, \bibinfo{person}{Yixuan Zhang}, \bibinfo{person}{Yiqiao Jin}, \bibinfo{person}{Kaijie Zhu}, \bibinfo{person}{Hao Chen}, {and} \bibinfo{person}{Xing Xie}.} \bibinfo{year}{2023}\natexlab{}.
\newblock \showarticletitle{Competeai: Understanding the competition behaviors in large language model-based agents}.
\newblock \bibinfo{journal}{\emph{arXiv preprint arXiv:2310.17512}} (\bibinfo{year}{2023}).
\newblock


\bibitem[Zheng et~al\mbox{.}(2024)]%
        {m27}
\bibfield{author}{\bibinfo{person}{Junhao Zheng}, \bibinfo{person}{Shengjie Qiu}, \bibinfo{person}{Chengming Shi}, {and} \bibinfo{person}{Qianli Ma}.} \bibinfo{year}{2024}\natexlab{}.
\newblock \showarticletitle{Towards Lifelong Learning of Large Language Models: A Survey}.
\newblock \bibinfo{journal}{\emph{arXiv preprint arXiv:2406.06391}} (\bibinfo{year}{2024}).
\newblock


\end{thebibliography}

\appendix

\end{document}